\documentclass[twocolumn]{aastex63}
\makeatletter
\newcommand*{\rom}[1]{\expandafter\@slowromancap\romannumeral #1@}
\makeatother

\usepackage[utf8]{inputenc}
\usepackage{chngpage}
\usepackage{float}
\usepackage{hyperref}
\usepackage{longtable}

\usepackage{graphicx}	
\usepackage{amsmath}	
\usepackage{amssymb}	
\usepackage[T1]{fontenc}

\shorttitle{Dichotomy in Long-Lived Radio Emission from TDEs}
\shortauthors{Christy et al.}
\graphicspath{{./}{figures/}}

\begin{document}
\title[Dichotomy in Long-Lived Radio Emission from TDEs]{Dichotomy in Long-Lived Radio Emission from Tidal Disruption Events AT 2020zso and AT 2021sdu: Multi-Component Outflows vs. Host Contamination}
\newcommand{\UA}{\affiliation{Steward Observatory, University of Arizona, 933 North Cherry Avenue, Tucson, AZ 85721-0065, USA}}

\author[0000-0003-0528-202X]{Collin~T.~Christy}
\altaffiliation{E-mail: collinchristy@arizona.edu}
\UA

\author[0000-0002-8297-2473]{Kate D. Alexander}
\UA

\author[0000-0003-1792-2338]{Tanmoy Laskar}
\affiliation{Department of Physics \& Astronomy, University of Utah, Salt Lake City, UT 84112, USA}
\affiliation{Department of Astrophysics/IMAPP, Radboud University, PO Box 9010, 6500 GL Nijmegen, The Netherlands}

\author[0000-0003-4537-3575]{Noah Franz}
\UA

\author[0000-0003-3441-8299]{Adelle~J.~Goodwin}
\affiliation{International Centre for Radio Astronomy Research, Curtin University, GPO Box U1987, Perth, WA 6845, Australia}

\author[0000-0002-0744-0047]{Jeniveve Pearson}
\UA

\author[0000-0002-9392-9681]{Edo Berger}
\affiliation{Center for Astrophysics $\vert$ Harvard \& Smithsonian, Cambridge, MA 02138, USA}

\author[0000-0001-7007-6295]{Yvette Cendes}
\affiliation{Center for Astrophysics $\vert$ Harvard \& Smithsonian, Cambridge, MA 02138, USA}

\author[0000-0002-7706-5668]{Ryan Chornock}
\affiliation{Department of Astronomy, University of California, Berkeley, CA 94720-3411, USA}
\affiliation{Berkeley Center for Multi-messenger Research on Astrophysical Transients and Outreach (Multi-RAPTOR), University of California, Berkeley, CA 94720-3411, USA}

\author[0000-0001-5126-6237]{Deanne Coppejans}
\affiliation{Department of Physics, University of Warwick, Gibbet Hill Road, Coventry CV4 7AL, UK}

\author[0000-0003-0307-9984]{Tarraneh Eftekhari}
\altaffiliation{NHFP Einstein Fellow}
\affiliation{Center for Interdisciplinary Exploration and Research in Astronomy (CIERA), Northwestern University, 1800 Sherman Avenue, Evanston, IL 60201, USA}

\author[0000-0003-4768-7586]{Raffaella Margutti}
\affiliation{Department of Astronomy, University of California, Berkeley, CA 94720-3411, USA}
\affiliation{Department of Physics, University of California, 366 Physics North MC 7300, Berkeley, CA 94720, USA}
\affiliation{Berkeley Center for Multi-messenger Research on Astrophysical Transients and Outreach (Multi-RAPTOR), University of California, Berkeley, CA 94720-3411, USA}

\author[0000-0003-3124-2814]{James C. A. Miller-Jones}
\affiliation{International Centre for Radio Astronomy Research, Curtin University, GPO Box U1987, Perth, WA 6845, Australia}

\author[0000-0001-6971-4851]{Melanie Krips}
\affiliation{Institut de Radioastronomie Millim\'{e}trique (IRAM), 300 Rue de la Piscine, 38400 Saint-Martin-d'H\`{e}res, France}

\author[0000-0003-2558-3102]{Enrico Ramirez-Ruiz}
\affiliation{Department of Astronomy and Astrophysics, UCO/Lick Observatory, University of California, 1156 High Street, Santa Cruz, CA 95064, USA}

\author[0000-0003-4102-380X]{David J. Sand}
\UA

\author[0000-0002-4912-2477]{Richard Saxton}
\affiliation{Operations Department, Telespazio UK for ESA, European Space Astronomy Centre, 28691 Villanueva de la Ca\~{n}ada,  Spain}

\author[0000-0002-4022-1874]{Manisha Shrestha}
\UA

\author[0000-0002-3859-8074]{Sjoert van Velzen}
\affiliation{Leiden Observatory, Leiden University, PO Box 9513, NL-2300 RA Leiden, the Netherlands}

\begin{abstract}
\indent We present a detailed radio study of the tidal disruption events (TDEs) AT~2020zso and AT~2021sdu. Both exhibit transient radio emission beginning shortly after optical discovery and persisting for several years. For AT~2020zso, we identify two distinct radio flares. The first arises soon after the optical peak, reaching a maximum $\sim1$ year post-discovery before fading. The second flare appears $\sim800$ days after discovery and results in the brief presence of two distinct components in the radio spectra, providing strong evidence for physically separate outflows. Both flares are consistent with non-relativistic outflows, with velocities $v\approx0.1-0.2c$ and energies $E\sim10^{49}$ erg, propagating through a Bondi-like circumnuclear medium. Our analysis supports a scenario in which the first outflow is accretion-driven, launched while the TDE disk is accreting at a relatively high Eddington fraction, whereas the second outflow is associated with a transition to an advection-dominated accretion flow. In contrast, the radio emission from AT~2021sdu is best explained by a slower ($v\approx0.03c$), less energetic outflow ($E\sim10^{48}$ erg), combined with diffuse, non-variable host emission that becomes dominant $\sim500$ days after discovery. Assuming free expansion, we infer an outflow launch date preceding the optical discovery date. This suggests that the outflow may originate from either the unbound stellar debris ejected during disruption or, alternatively, from a decelerating outflow. Our findings demonstrate the diversity of outflow properties in TDEs and highlight the observational challenges of interpreting late-time radio variability in the presence of host galaxy contamination.
\end{abstract}


\section{Introduction}
Stars that venture too close to supermassive black holes (SMBHs) experience tidal forces that can overpower their self-gravity, ultimately tearing them apart \citep{Rees1998,Evans_1989,Guillochon_2013}. This process results in a luminous transient in which roughly half of the stellar material accretes onto the black hole, while the rest becomes gravitationally unbound and is ejected into the surrounding medium; such phenomena are known as tidal disruption events (TDEs). These transients can produce detectable emission across the entire electromagnetic spectrum. 

TDEs are often discovered through their X-ray, UV, and/or optical emission (see, however, \citealt{ARP_2018,anderson_2020,jiang_2021,Masterson_2024,dykaar_2024,Somalwar_2025}). The optical/UV emission is thought to trace reprocessed radiation from a newly formed accretion flow (via an optically thick envelope), shocks produced by self-intersections of the bound debris streams, or a combination of these mechanisms (see \citealt{vanVelzen2020} for a review). Although models differ on the exact physical processes underlying this emission, they generally agree on the presence of energetic outflows as a common feature of TDEs (\citealt{alexander2020}). Radio observations in particular offer a complementary probe of these outflows as they interact with the circumnuclear environment.

It has become increasingly clear that TDEs can produce radio emission on a variety of timescales. In rare cases ($\lesssim1\%$), TDEs promptly launch powerful relativistic jets that generate luminous radio emission ($\nu L_{\nu,\text{radio}}\gtrsim10^{40}$erg/s) detectable from days to years post-disruption (\citealt{Giannios_2011}; \citealt{Zauderer_2011}; \citealt{De_Colle_2012}; \citealt{Andreoni_2022}). More commonly, TDEs produce lower-luminosity ($\nu L_{\nu,\text{radio}}\lesssim10^{40}$erg/s) radio emission on timescales of months consistent with slower outflows interacting with the circumnuclear medium (CNM; e.g., \citealt{alexander_2016,at2020opy,Christy_2024}). However, more than half of known TDEs are not detected in the radio during the first $\sim6$ months following the disruption \citep{alexander2020}. 

Recent radio monitoring has revealed that, for some TDEs ($\sim40\%$), radio emission emerges years after the initial disruption of the star (e.g., \citealt{horesh_15oi,AT2018hyz,Cendes_2024}). In a few cases, these late-time flares are indicative of renewed energy injection or multiple ejection episodes, often without any changes in the UV/optical emission (\citealt{horesh_15oi}; \citealt{Hajela_2024}; \citealt{Goodwin_2025}). The origin of these delayed flares remains debated, with proposed explanations ranging from emerging jets (e.g., \citealt{Teboul_2023,Lu_2024,Sfaradi_2024}), changes in the accretion flow (e.g., \citealt{sfaradi_2022,Piro_2024, Alexander_2025}), or environmental factors (e.g., \citealt{ Mou_2022,Matsumoto_2024,Zhuang_2024}).

Motivated by the growing number of TDEs with late-time radio flares and the need for detailed radio monitoring at both early and late times to distinguish among the above possible scenarios, we present a detailed study of two events: AT 2020zso and AT 2021sdu. In both cases, the radio spectra evolve in a manner inconsistent with a single emission component. Instead, the data suggest that the radio emission is intrinsically multi-component in nature.

We organize this paper as follows. In \S \ref{sec:obs}, we describe our radio observations and data reduction methods. In Section \S \ref{sec:host_modeling}, we examine the properties of the host galaxies. In Section \S \ref{sec:lcmodeling}, we discuss the radio light curves. In Section \S \ref{sec:modeling}, we present a model for the radio spectral energy distributions. In Section \S \ref{sec:analysis}, we infer the physical properties of the outflows and we discuss our findings in Section \S \ref{sec:disc}. In Section \S \ref{sec:conc}, we summarize our conclusions.

\section{Observations} \label{sec:obs}
The transient event AT 2020zso was first discovered by the Zwicky Transient Facility (ZTF; \citealt{Bellm_2019}) on 2020 Nov 12 UTC \citep{20zso_TNS_discovery} and was subsequently classified as a TDE based on UV/optical photometric and spectroscopic observations \citep{2020zso_classification}. \citet{Wevers_2022} later performed a detailed analysis of this object, focusing on the evolution of its UV/optical emission. They show that the host galaxy, SDSS J222217.13-071558.9 ($z = 0.0563$), contains an active galactic nucleus (AGN), and suggest that this event is likely a partial stellar disruption due to evidence for a low impact parameter ($\beta<0.9$) and low stellar debris mass ($\sim0.1M_\odot$). They also identified transient, asymmetric double-peaked line profiles in He~\rom{2} and H$\alpha$, consistent with a highly elliptical, nearly edge-on accretion disk. Separately, AT 2021sdu was discovered by ZTF on 2021 Jul 5 UTC and soon after identified as a TDE \citep{Chu_2021}. Analysis of AT 2021sdu as part of the ZTF sample presented in \citet{Yao_2023} found no unusual spectroscopic or light curve features in comparison to other optical TDEs.

After their initial classifications, we began 3+ years of radio monitoring for the TDEs AT 2020zso and AT 2021sdu. Additionally, we obtained late-time optical spectra of each TDE's host galaxy. For all observations, we define $\delta t$ as the time relative to the optical discovery dates recorded by the Transient Name Server (TNS\footnote{\href{https://www.wis-tns.org/}{https://www.wis-tns.org/}}): MJD 59165.7 (AT 2020zso; \citealt{20zso_TNS_discovery} and MJD 59400.9 (AT 2021sdu; \citealt{21sdu_TNS_discovery}). For our analysis, we assume luminosity distances of $D_L = 251$~Mpc ($z = 0.0563$) for AT~2020zso and $D_L = 264$~Mpc ($z = 0.059$) for AT~2021sdu, using a flat $\Lambda$CDM cosmology with $H_0 = 70~\text{km~s}^{-1}~\text{Mpc}^{-1}$, $\Omega_m = 0.3$, and $\Omega_\Lambda = 0.7$.

\subsection{VLA}\label{subsec:VLA}
We obtained radio observations for both sources with the Karl G. Jansky Very Large Array (VLA) under the program 20B-377 (PI: K. Alexander). Over the course of our observations, we utilized the observing bands: $L$ (1.5 GHz), $S$ (3 GHz), $C$ (6 GHz), $X$ (10 GHz), $Ku$ (15 GHz), $K$ (22 GHz), $Ka$ (33 GHz), and $Q$ (44 GHz). We performed wide-band continuum observations using 8-bit samplers at $L-$band and $S$-band and 3-bit samplers at all other frequencies to maximize sensitivity. For AT 2020zso, we observed in $L/S/C/X/Ku$ bands, while for AT 2021sdu we obtained broader coverage including $L/S/C/X/Ku/K/Ka/Q$ bands. Details of the observing setup are as follows:
\begin{itemize}
    \item \textbf{AT 2020zso}: We used 0137+331=3C48 as the bandpass and flux calibrator for all observations and frequencies. For VLA configurations B, C, and D, we employed J2229-0832 as the gain calibrator for all frequencies. For observations in the VLA A configuration, J2246-1206 was used as the gain calibrator for $L/S/C-$bands and J2229-0832 was used for all remaining frequencies.

    \item \textbf{AT 2021sdu}: We used 0542+498=3C147 as the bandpass and flux calibrator for all observations and frequencies. We employed J0105+4819 as the gain calibrator for all observations at $C/X/Ku/K/Ka/Q-$bands. For $L-$ and $S-$band, we used J0136+4751 as the gain calibrator.
\end{itemize}
The data were processed using standard reduction procedures in the Common Astronomy Software Application (CASA; \citealt{Casa,CASA_team}). We imaged the data from each receiver band using the CASA task \verb'tclean'. When the targets were sufficiently bright, we further split the data and imaged each subband. We obtained flux densities and their uncertainties by using the CASA task \verb'imfit' to fit to the source an elliptical Gaussian fixed to the size of the synthesized beam. The final flux density measurements from our VLA analysis are summarized in Tables \ref{tab:data_zso} and \ref{tab:data_sdu}.

We note that the uncertainties derived from our fitting procedure are statistical errors only. When modeling, we include an additional uncertainty of 5\% of the source flux density to account for the known absolute flux density scale calibration accuracy\footnote{For more details about uncertainties in the VLA flux density scale, see: \url{https://science.nrao.edu/facilities/vla/docs/manuals/oss/performance/fdscale}.}.

\subsection{GMRT}
We obtained additional observations of AT 2020zso and AT 2021sdu with the Giant Metrewave Radio Telescope (GMRT) in Bands 4 and 5 (centered at 0.65 and 1.26 GHz, respectively) using the full 400 MHz bandwidth provided by the GMRT Wideband Backend (GWB) system. Below, we outline the details of the GMRT observations presented in this work:

\begin{itemize}
    \item \textbf{AT 2020zso}: We employed 3C48 as the bandpass and flux calibrator and J2246-1206 as the gain calibrator for all observations. These observations were conducted under the program 43\_046 (PI: A. Goodwin).
    \item \textbf{AT 2021sdu}: We employed 3C147 as the bandpass and flux calibrator and J0136+4751 as the gain calibrator for all observations. These observations were conducted under the programs 44\_022 and 45\_123 (PI: A. Goodwin).
\end{itemize}
We followed the same calibration and imaging procedures described in Section~\ref{subsec:VLA}. For AT 2020zso, we include in our analysis the Band 4 and Band 5 upper limits reported in ATel~\#14828 \citep{Roy_2021}, obtained on 2021 May 30 UT and 2021 Jun 8 UT ($\delta t = 198, 208$ d), respectively.

\subsection{NOEMA}
We obtained radio observations of AT 2021sdu with the Northern Extended Millimeter Array (NOEMA) under the programs E21AA and S22BS (PI: T. Laskar). We observed using the 3 mm receiver centered at 88.5 GHz. For calibration, we typically used 3C84 as the bandpass calibrator, MWC349 as the flux calibrator, and J0136+478 as the gain calibrator.\footnote{This was the standard calibrator setup, except on 2022 January 26, when LKHA101 was used as the flux calibrator, and on 2022 September 30, when 3C454.3 was used as the bandpass calibrator.} The data were reduced in GILDAS \footnote{\href{https://www.iram.fr/IRAMFR/GILDAS/}{https://www.iram.fr/IRAMFR/GILDAS/}} following standard calibration procedures. When modeling, we include an additional 5\% systematic error to capture the uncertainty on the absolute flux density calibration.

\subsection{Bok 90''}
We obtained low-resolution optical spectra of the host galaxies of AT 2020zso and AT 2021sdu with the Boller \& Chivens (B\&C) spectrograph on the University of Arizona's 2.3~m Bok telescope at Kitt Peak National Observatory (KPNO). The observations used a $1.5^{\prime\prime}$ slit width and a grating centered at $6690$~\AA. The data were reduced using standard techniques in IRAF. For AT 2020zso and AT 2021sdu, the spectra were obtained at $\delta t \approx 1469$ days and $\delta t \approx 1234$ days, respectively. We show the host spectra in Figure~\ref{fig:hostspec}.

\section{Host Galaxy Properties}\label{sec:host_modeling}

\begin{figure*}[ht!]
    \centering
    \includegraphics[width = \textwidth]{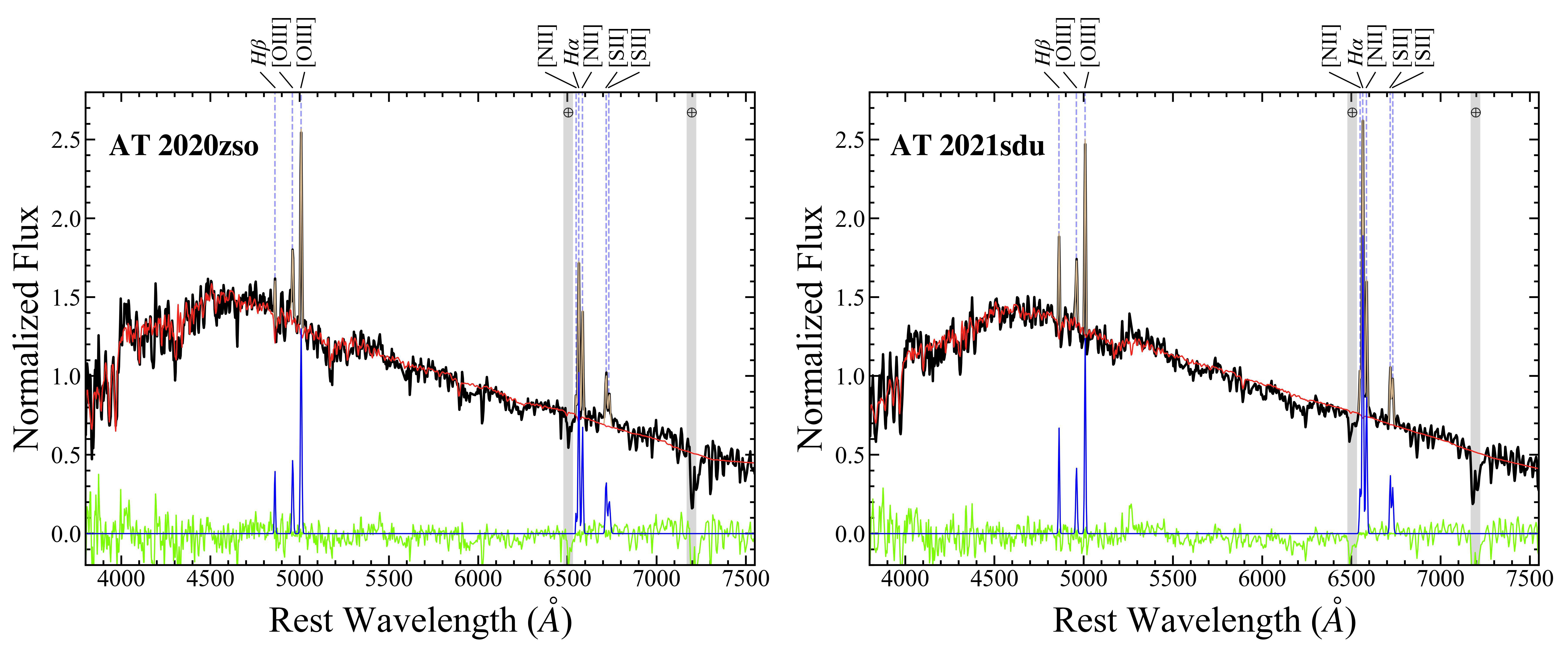}
    \caption{Late-time optical spectra (black lines) of the host galaxies of AT 2020zso (SDSS J222217.13-071558.9; \textit{Left}) and AT 2021sdu (WISEA J011123.92+503429.7;\textit{Right}) obtained at $\delta t \approx 1469, 1234$ days after discovery. The red lines show the best-fit \texttt{ppxf} models for the stellar continua, while the tan curves represent fits to the gas emission lines. The green lines show the residuals, and the blue curves indicate the continuum-subtracted emission lines. Telluric bands are shown in light gray.}
    \label{fig:hostspec}
\end{figure*}

\begin{figure}
    \centering
    \includegraphics[width = 0.5\textwidth]{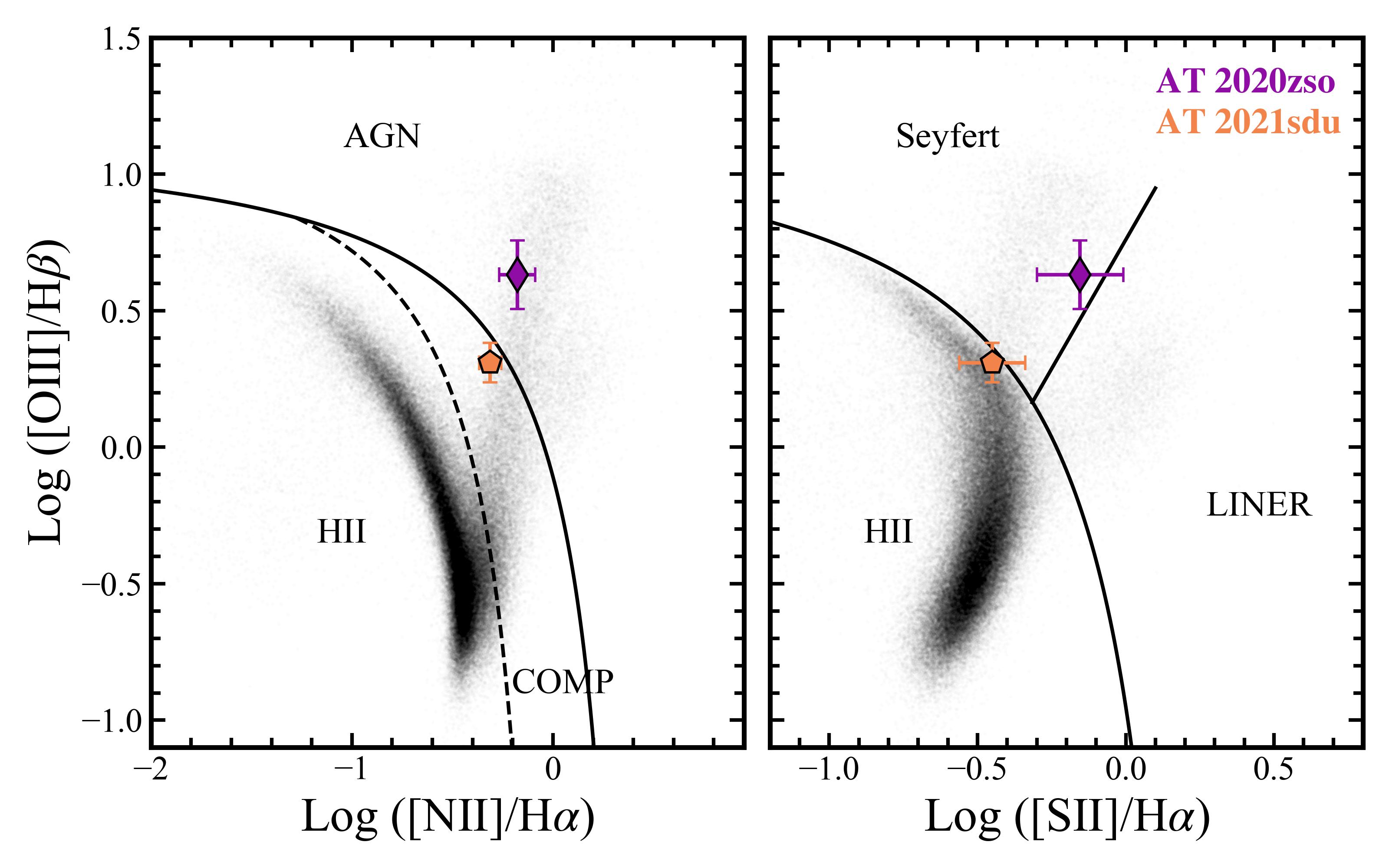}
    \caption{BPT diagrams showing the hosts of AT 2020zso and AT 2021sdu. The solid and dashed dividing lines are defined in \citet{Kauffmann_2003} and \citet{Kewley_2006}. (\textit{Left}) Galaxies are classified as star-forming (H~\rom{2}), AGN, or composite (COMP; H~\rom{2} + AGN). (\textit{Right}) Galaxies are separated into star-forming (H~\rom{2}), Seyfert AGN (high-ionization), and LINERs (low-ionization narrow emission-line regions). The galaxies shown in gray are from the Sloan Digital Sky Survey (SDSS) 7th data release \citep{oh_2011}. The narrow line ratios confirm the host of AT 2020zso is likely an AGN in agreement with the findings of \citet{Wevers_2022}. We find that line ratios for AT 2021sdu suggest a combination of star formation and AGN activity.}
    \label{fig:bpt}
\end{figure}

The host galaxy of AT 2020zso (SDSS J222217.13-071558.9) has been previously investigated by \citet{Wevers_2022}, who identified it as a Seyfert galaxy based on the presence of high-ionization narrow emission lines. Due to the lower resolution of the B\&C spectrograph, we did not recover all of the emission features seen by \citet{Wevers_2022}. However, we were able to identify H$\beta$, the [O~\rom{3}] $\lambda\lambda$4959, 5007 doublet, the [N~\rom{2}] $\lambda\lambda$6548, 6584 doublet, H$\alpha$, and the [S~\rom{2}] $\lambda\lambda$6717, 6731 doublet. To extract the emission line fluxes, we use the penalized pixel-fitting (\verb'ppxf') software \citep{Cappellari_Emsellem_2004,Cappellari_2017,Cappellari_2023} to fit the extinction corrected B\&C spectrum to stellar population models using the MILES library (FWHM = 2.5~\AA; \citealt{MILES,Vazdekis_2016}). After subtracting the underlying stellar continuum, we fit for the narrow emission lines (see Figure \ref{fig:hostspec}). Using these features, we compute the line ratios and display them on a Baldwin-Phillips-Terlevich (BPT) diagram \citep{BPT}, as shown in Figure \ref{fig:bpt}. Our results are consistent with \citet{Wevers_2022} in that the host galaxy of AT 2020zso is likely a Seyfert type 2. 

Similarly, we were able to identify the same handful of narrow emission line features for the host galaxy of AT 2021sdu (WISEA J011123.92+503429.7). As shown in Figure \ref{fig:bpt}, the line ratios for the host fall within the composite region, consistent with a combination of both star formation and an AGN. Alternately the elevated high-ionization lines may instead indicate that AT2021sdu's host galaxy has both star formation and an enhanced TDE rate, as the cumulative ionizing output from repeated TDE disks can strengthen line emission beyond that produced by star formation alone \citep{Mummery_2025}.

\subsection{Radio Emission from the Host of AT 2021sdu}

\begin{figure}[ht!]
    \centering
    \includegraphics[width = 0.475\textwidth]{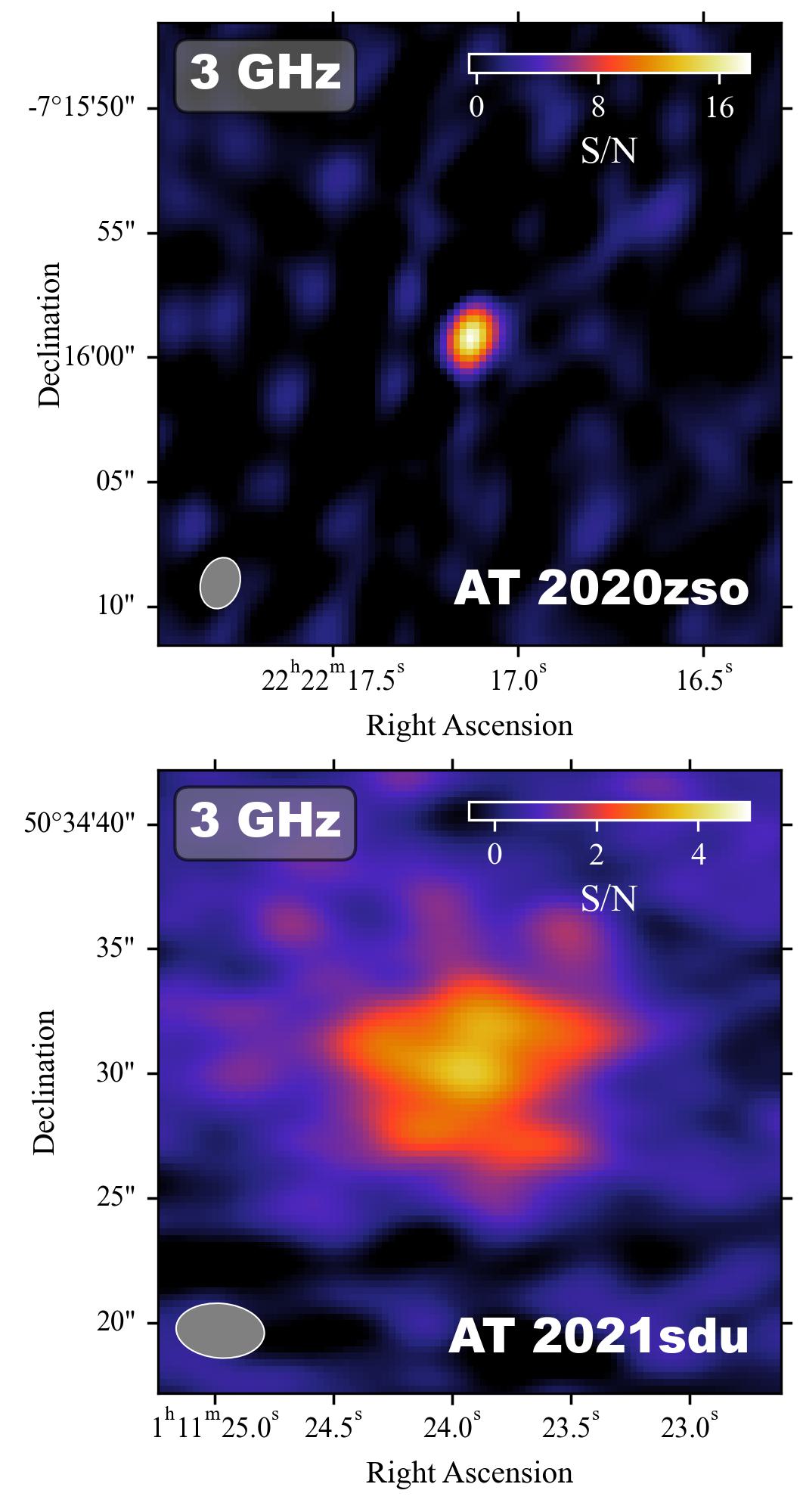}
    \caption{Late-time VLA 3 GHz images of AT 2020zso and AT 2021sdu taken on 2024 May 9 ($\delta t\approx1275$ d) and 2024 May 10 ($\delta t\approx 1040$ d), respectively. We see that AT 2021sdu exhibits diffuse radio emission at low frequencies, while AT 2020zso appears as an unresolved point source. The color bar in the upper right indicates the signal‑to‑noise ratio, and the synthesized beam is shown in the bottom left for reference.}
    \label{fig:radioimages}
\end{figure}

Consistent with the above picture, our radio observations also reveal ongoing star formation. In the most extended VLA configurations, which provide the highest angular resolution, the central radio source appears resolved rather than point-like at low frequencies ($\nu\lesssim 3$ GHz; see Figure \ref{fig:radioimages}). Additionally, we find no evidence of temporal variability in either our GMRT Band 4 (0.65 GHz observer frame) or Band 5 (1.26 GHz observer frame) observations from $\delta t = 655 - 965 ~\rm{d}$. We interpret this as evidence for a persistent host galaxy component that dominates emission at frequencies $\nu \lesssim 3$GHz. This component becomes more prominent at late times, around $\sim 1$ year post-optical discovery, when the transient radio emission from the TDE weakens and we begin observing at lower frequencies. Once the transient component fades, we find that any remaining observed temporal variability can be interpreted as resolution effects from observing diffuse host emission at different spatial scales.

To examine the transient emission, we need to properly account for the host contribution in all epochs. However, since the VLA data span multiple array configurations, the amount of resolved host emission will vary across the observations. Given that the VLA data for AT 2021sdu were taken in the A, B, and C configurations, we need a configuration-dependent spectral model of the host emission. Since the GMRT maintains a fixed configuration and the GMRT photometry was non-variable, we use the latter as the basis for deriving the host models.

We model the flux density of the host as a power-law defined as:

\begin{equation}
    F_{\nu,\rm Host} =F_{0} \left(\frac{\nu}{\rm{GHz}}\right)^{-\alpha_0}
    \label{eq:host_model}
\end{equation} 

\noindent  To constrain the configuration-dependent model parameters, we re-imaged the GMRT data using the VLA-B ($0.21~\rm{km} < uv\text{-range} < 11.1~\rm{km}$) and VLA-C ($0.035~\rm{km} < uv\text{-range} < 3.4~\rm{km}$) $uv$-range constraints. We then fit this model separately to each  $uv$-range constraint, ensuring that they share a common spectral index $\alpha_0$. We find the best fit values $F_{0,B} = 323_{-26}^{+26}\rm{\mu Jy}$, $F_{0,C} = 606_{-54}^{+54}\rm{\mu Jy}$, and $\alpha_0 = 1.3_{-0.2}^{+0.2}$, where the subscript on $F_0$ indicates the corresponding VLA configuration. These values define the configuration-dependent host models used in this analysis. For data taken in VLA-A configuration, we adopt the VLA-B host model. The VLA-A $L$-band data were imaged with the VLA-B $uv$-range constraints to ensure self-consistency; applying the same restriction at higher frequencies did not significantly affect the derived point-source flux densities at the phase center.

When imaging the GMRT data for AT 2021sdu using the VLA-C configuration's $uv$-range constraints, the Band 5 beam is large enough to recover the full extended flux which corresponds to $L_{1.4\rm{GHz}}=3.0_{-0.3}^{+0.3}\times10^{28}~\rm{erg}~\rm{s^{-1}~\rm{Hz^{-1}}}$. Using the radio luminosity to star formation rate (SFR) calibration from \citet{murphy_2011}, we find that this host component implies a SFR $\approx 1.87_{-0.19}^{+0.20} \ M_\odot \rm{yr}^{-1}$. Separately, the measured $H_\alpha$ line luminosity, $L_{\rm{H\alpha}}\approx1.2\times10^{41} \rm{erg} \ \rm{s}^{-1}$, implies a lower limit on the SFR of $\gtrsim 0.67 \ M_\odot \rm{yr}^{-1}$ \citep{murphy_2011}. The lower limit arises from uncertainties in the fraction of ionizing photons that are absorbed by dust. Given the agreement between these independent tracers of star formation, we conclude that the host radio emission in AT 2021sdu is likely dominated by star formation. However, as noted above the presence of stronger than expected high-ionization lines (e.g., [O~\rom{3}] $\lambda\lambda$4959, 5007 and [N~\rom{2}] $\lambda\lambda$6548, 6584) prevents us from ruling out a contribution from a weak compact AGN, although this could also arise naturally if the TDE rate is enhanced in this galaxy (see \citealt{Mummery_2025}). 

Subtracting the host component (Equation \eqref{eq:host_model}) from the data yields temporally evolving radio spectra consistent with a radio transient associated with AT 2021sdu during the first 6 epochs ($\delta t\sim68 - 461$ d). For data taken after $\delta t \gtrsim 461$ days, we find that our radio data are consistent with that of the host galaxy alone. Our subsequent analysis of AT 2021sdu will focus on interpreting the transient component remaining after host subtraction. 

\section{Radio Light Curve Modeling}\label{sec:lcmodeling}

\begin{figure*}
    \centering
    \includegraphics[width = \textwidth]{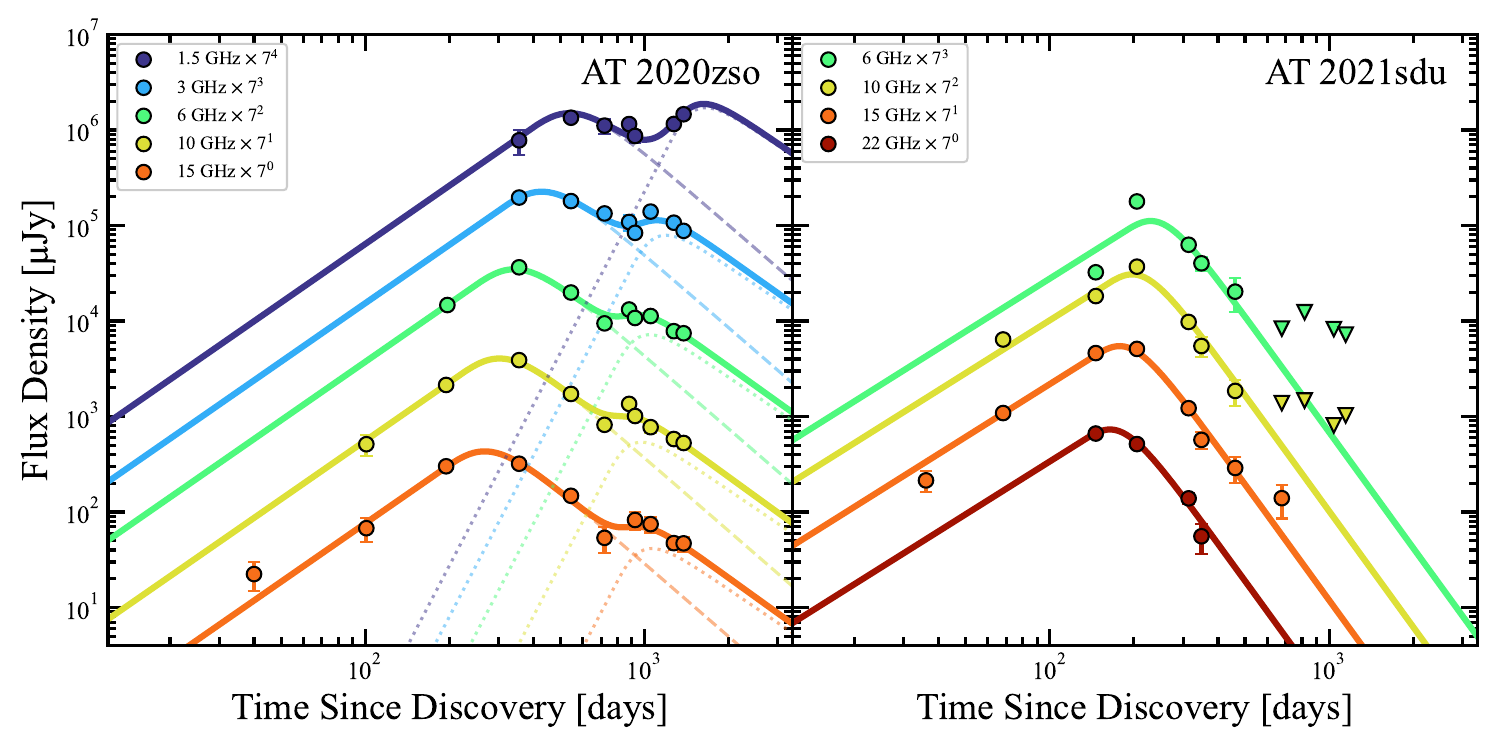}
    \caption{Multi-frequency radio light curves for AT 2020zso (\textit{Left}) and AT 2021sdu (\textit{Right}). The solid line for AT 2020zso represents the best-fit solution for a two-component outflow at each frequency, while the dashed and dotted lines correspond to the best-fit single broken power-law solutions for the prompt and delayed radio flares, respectively. For AT 2021sdu, we show the host-subtracted flux densities. Solid lines represent the best-fit broken power-law solutions at each frequency.}
    \label{fig:lightcurves}
\end{figure*}

We show the radio light curves of AT 2020zso and AT 2021sdu for the best-sampled frequencies in Figure \ref{fig:lightcurves}. Most notably, we find that AT 2020zso displays two distinct peaks in its radio light curve, similar to AT 2019dsg and AT 2020vwl (\citealt{Cendes_2024}; \citealt{Goodwin_2025}). We find no evidence for a second peak in AT 2021sdu's host-subtracted radio light curve. 

The radio emission from both events rises and decays in a frequency-dependent manner, with the flux at higher frequencies decaying first. To capture the rise and decay rates of the radio emission at each frequency, we modeled the radio light curves as broken power-laws following the prescription from \citet{1998chevalier},

\begin{align}
    F_\nu(t) &= 1.582 F_\nu(t_c) \left(\frac{t}{t_c}\right)^a \nonumber\\
    &\times\left(1-\text{exp}\left[-\left(\frac{t}{t_c}\right)^{-(a+b)}\right]\right)
    \label{eq:chev_lc}
\end{align}

\noindent where $a$ and $b$ are the power law exponents of the rising and decaying segments, respectively, and $F_\nu(t_c)$ is the flux density at the break time $t_c$. We jointly fit the light curves at different frequencies to have the same power-law rise and decay slopes. This restriction is a valid assumption as the slopes primarily depend on the expansion rate of the emitting region and the power-law index of the distribution of relativistic electron energies, both of which should be constant given the expected evolution of a single forward shock in a power-law density profile \citep{1998chevalier}. 

For AT 2021sdu, we fit the data using Equation \eqref{eq:chev_lc}, which models the emission as a single broken power law. In contrast, the radio light curve of AT 2020zso shows evidence for two peaks, so we adopt a modified model with two broken power-law components:
\begin{align}
    F_\nu(t) &= F_\nu(t,t_{c,1},a_1,b_1) \nonumber \\
    &+ F_\nu(t,t_{c,2},a_2,b_2) \quad \text{where } t_{c,1}<t_{c,2} 
\end{align}

To determine the best-fitting model parameters, we fit each broken power law model using the Markov Chain Monte Carlo (MCMC) module \verb'emcee' \citep{emcee} in Python, assuming a Gaussian error model for the measurements. We adopt uniform priors on the model parameters $a$, $b$, $t_c$, and $F_\nu(t_c)$, where $a,b \in [1, 10]$, $\text{log}_{10}\left(F_\nu(t_c)/\mu\text{Jy}\right) \in [1,3.3]$, and $\text{log}_{10}\left(t_c/\text{days}\right) \in [1,3.3]$. When fitting for AT 2020zso, we implement an additional prior $t_{c,1}<t_{c,2}$. When fitting for two components, we allow the rise and decay rates $a_1$, $a_2$ and $b_1$, $b_2$ to differ, given that they may not be driven by the same forward shock component. We report the best-fit model parameters in Table \ref{tab:lc_model_params} and display these fits in Figure \ref{fig:lightcurves}.

\begin{table}
    \centering
    \resizebox{0.5\textwidth}{!}{%
        \makebox[0.5\textwidth][r]{ 
            \begin{tabular}{l|ccccc}
            \hline\hline
            Object & $\nu_\text{obs}$ & $F_\nu(t_{c,1})$& $t_{c,1}$ & $F_\nu(t_{c,2})$& $t_{c,2}$ \\
                   & (GHz) & ($\mu$Jy) & (days) & ($\mu$Jy)& (days) \\
            \hline
            {\small \textbf{\small AT 2020zso}}& 1.5 & $607_{-28}^{+34}$ & $584_{-20}^{+11}$ & $>452$          & $>1246$   \\
            $a_1=2.0_{-0.2}^{+0.2}$         & 3   & $641_{-35}^{+32}$ & $464_{-13}^{+9} $ & $222_{-19}^{+18}$& $1112_{-39}^{+49}$ \\
            $b_1=2.5_{-0.2}^{+0.2}$          & 6   & $692_{-24}^{+22}$ & $371_{-10}^{+9} $  & $141_{-11}^{+7}$ & $971_{-27}^{+24}$ \\
            $a_2=5.6_{-1.8}^{+1.8}$         & 10  & $564_{-18}^{+16}$ & $328_{-5}^{+5}  $ & $74_{-7}^{+5}$   & $910_{-33}^{+27}$ \\
            $b_2=2.0_{-0.5}^{+0.5}$         & 15  & $421_{-14}^{+25}$ & $291_{-7}^{+6}  $ & $40_{-5}^{+5}$   & $991_{-80}^{+120}$ \\
            \hline 
            {\small \textbf{AT 2021sdu}}& 6  & $296_{-14}^{+14}$  & $258_{-5}^{+5}$&-&- \\
            $a_1=1.8_{-0.1}^{+0.1}$    & 10 & $572_{-17}^{+25}$  & $222_{-3}^{+4}$&-&- \\
            $b_1=4.0_{-0.3}^{+0.3}$    & 15 & $710_{-35}^{+26}$  & $200_{-3}^{+3}$&-&- \\
                & 22 & $665_{-53}^{+58}$& $185_{-5}^{+10}$&-&- \\
            \hline\hline
            \end{tabular}
        }
    }
    \caption{MCMC results for the light curve fitting procedures outlined in Section \S \ref{sec:lcmodeling}.}
    \label{tab:lc_model_params}
\end{table} 

\section{Synchrotron Emission Modeling}\label{sec:modeling}
Outflows from TDEs are expected to produce synchrotron emission as the ejected material interacts with and expands into the surrounding medium. In this process, the outflow drives a shock into this medium, accelerating free electrons into a power-law distribution of energies described by $N(\gamma)\propto\gamma^{-p}$ for $\gamma>\gamma_m$, where $\gamma$ is the electron Lorentz factor, $p$ is the power-law index of the energy distribution, and $\gamma_m$ is the minimum Lorentz factor of the shocked electrons. The resulting synchrotron spectral energy distribution (SED) typically follows a broken power-law structure, characterized by distinct break frequencies ($\nu_m, \nu_a, \nu_c$) and an overall normalization factor $F_p$ \citep{Granot_2002}. Here, $\nu_m$ is the synchrotron frequency of the electrons with the minimal energy of the power law, $\nu_a$ is the synchrotron self-absorption frequency, and $\nu_c$ is the synchrotron cooling frequency. For a single break, the spectrum can be modeled as follows:

\begin{equation}\label{eq:sed}
    F_\nu(\nu) =  F_{\nu_b}\left[\left(\frac{\nu}{\nu_b}\right)^{-s\beta_1}+\left(\frac{\nu}{\nu_b}\right)^{-s\beta_2}\right]^{-1/s}
\end{equation}

\noindent where $\nu_b$ is the break frequency, $F_{\nu_b}$ is the extrapolated flux density at $\nu_b$, $\beta_1$, $\beta_2$ are the spectral slopes below and above $\nu_b$, and $s$ determines the smoothing. We find that the data for AT 2020zso and AT 2021sdu are consistent with a singly-broken power law with $\nu_b$ corresponding to the synchrotron self-absorption frequency $\nu_a$. Additionally, we fix $s=1$ for all epochs.

\subsection{AT 2020zso: A two component synchrotron spectrum}\label{subsec:AT 2020zso_model}
For AT 2020zso, we assume the spectral break frequencies are ordered as $\nu_m < \nu_a < \nu_c$. With this arrangement, it follows from \cite{Granot_2002} that $\beta_1=5/2$ and $\beta_2=(1-p)/2$.

\begin{figure*}[t!]
    \centering
    \includegraphics[width = \textwidth]{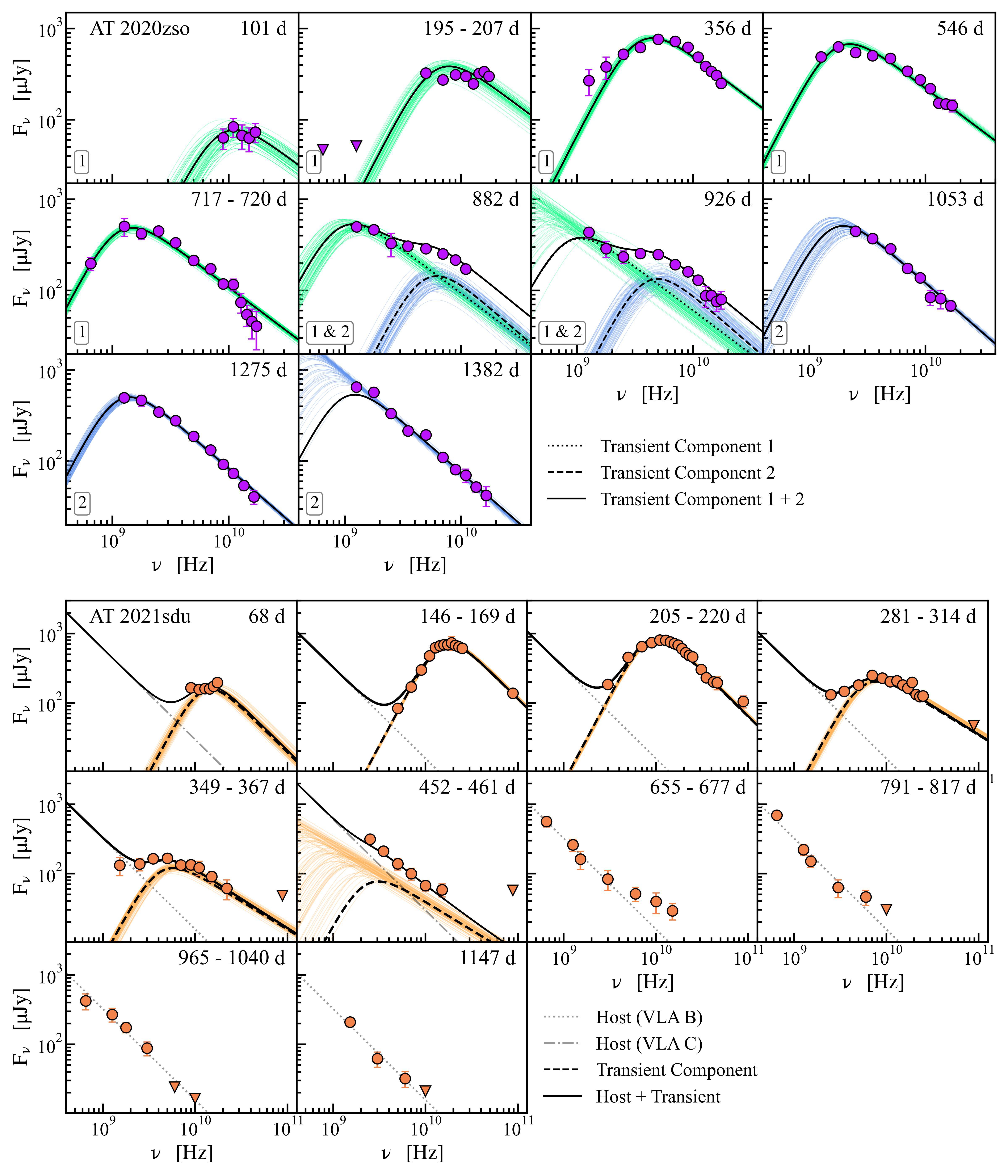}
    \caption{Radio spectra of AT 2020zso (\textit{Top}) and AT 2021sdu (\textit{Bottom}) at selected epochs. Observed data (solid points) and 100 MCMC posterior samples are color-coded: green/blue for AT 2020zso's two outflows; orange for AT 2021sdu. The total emission (solid black line) combines components: dotted and dashed lines represent AT 2020zso's first and second outflows, respectively; dashed and dash-dot/dotted lines denote AT 2021sdu's transient and configuration-dependent host emission. Panels indicate time since discovery (days) for both objects and include the contributing outflow components for AT 2020zso. For the epochs of AT 2020zso Outflow 1 ($\delta t=926$ days), Outflow 2 ($\delta t=1382$ days), and AT 2021sdu ($\delta t=461$ days), we do not capture the spectral peak ($\nu_p,F_{\nu,p}$); instead, we adopt an upper limit on $\nu_p$ and a lower limit on $F_{\nu,p}$ at these epochs.}
    \label{fig:dual_sed}
\end{figure*}

As discussed in Section \ref{sec:lcmodeling}, there is evidence for a two-component outflow in AT 2020zso. For brevity, we will refer to the initial component as Outflow $1$ and the latter as Outflow $2$.  As shown in Figure \ref{fig:lightcurves}, the radio light curve transitions from being dominated by emission from Outflow $1$ to emission from Outflow $2$ around $\delta t\sim900$ days. During this transition period, we observe two distinct peaks in the radio spectra at $\delta t =$ 882 days and 926 days (see Figure \ref{fig:dual_sed}). In particular, epoch $\delta t=926$ days most clearly captures the two peaks, which cannot be explained by any ordering of spectral breaks for a single synchrotron-emitting region \citep{Granot_2002}. We therefore interpret the spectral shape during these epochs as comprising two components, each characterized by a broken power-law with a well-defined peak. We consider the model spectrum for these cases as the sum of two components:
\begin{equation}
    F_\nu(\nu) =  F_{\nu,1}(\nu,\nu_a,F_{\nu_a},p_1) + F_{\nu,2}(\nu,\nu_a,F_{\nu_a},p_2)
\end{equation}
\noindent where $F_{\nu,1}(\nu)$ and $F_{\nu,2}(\nu)$ are each described by Equation \eqref{eq:sed}. For epochs prior to $\delta t=882$ days, we only fit for $F_{\nu,1}(\nu)$, while for epochs following $\delta t=926$ days, we only fit for $F_{\nu,2}(\nu)$. We assume that the contribution from the first outflow is subdominant in the later epochs and fit only a single component for simplicity. If we instead account for additional flux from the first outflow, assuming its peak frequency continues to evolve to lower frequencies at the same rate, its contribution to the overall spectrum remains minimal and does not affect the conclusions presented in this work.

\subsection{AT 2021sdu}\label{subsec:AT 2021sdu_model}
\begin{figure}
    \centering
    \includegraphics[width = 0.5\textwidth]{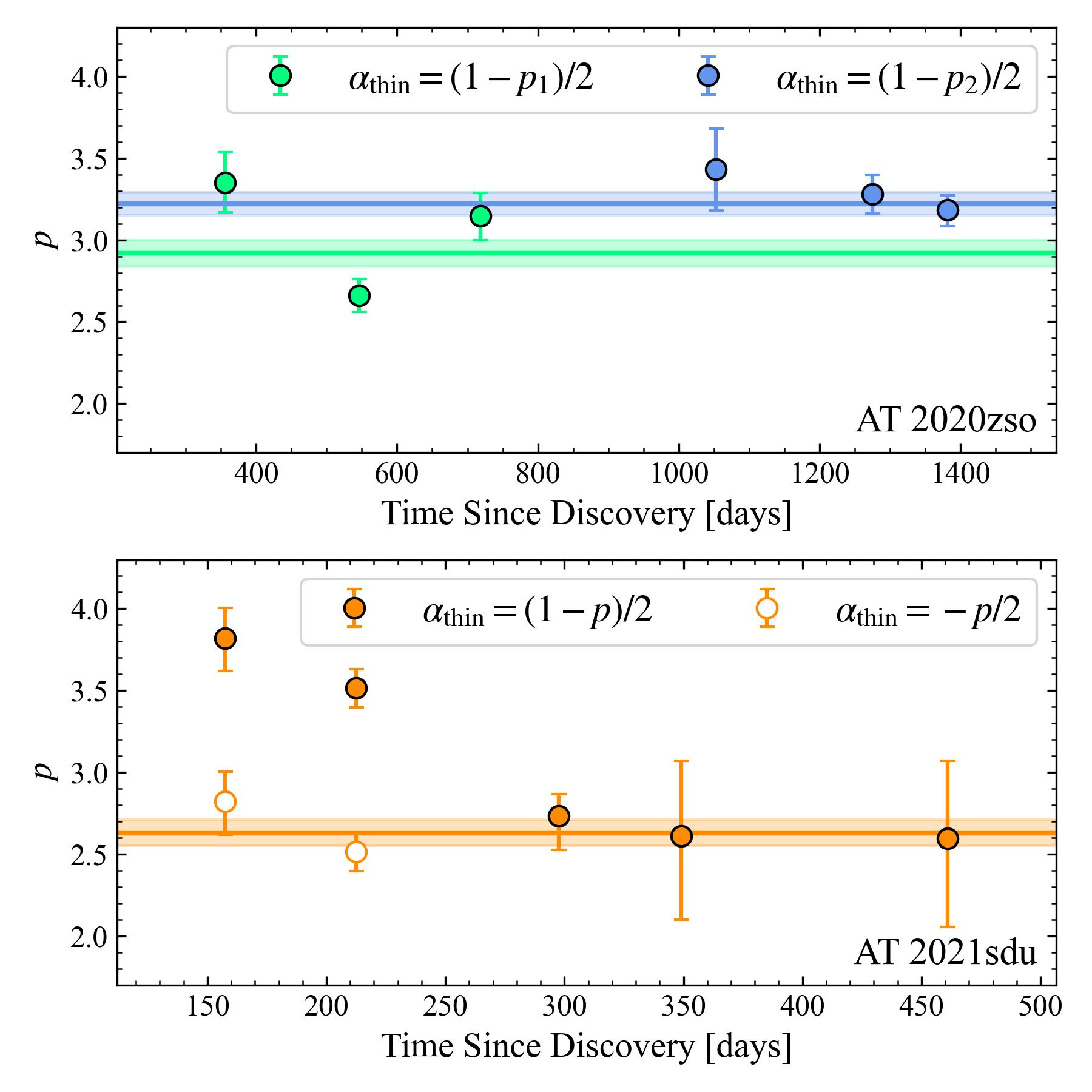}
    \caption{Electron power-law index $p$ over time for AT 2020zso (\textit{Top}) and AT 2021sdu (\textit{Bottom}) under different assumptions of $\alpha_{\rm{thin}}$, where $F_\nu \propto \nu^{\alpha_{\rm{thin}}}$ describes the optically thin emission. The solid lines and shaded regions show the weighted average $p$ and the $1\sigma$ uncertainty, respectively. For AT 2020zso, the subscript of $p$ denotes the numbered outflow dominating the spectra at that time.}
    \label{fig:spec_index}
\end{figure}

After subtracting the host model for AT 2021sdu (see Section \S \ref{sec:host_modeling}), the residual spectra exhibit a single peak, which we interpret as $\nu_a$. All of our radio spectra are well described by a single broken power-law, with no evidence for a second spectral break within the observed frequency range, suggesting that $\nu_m$ and $\nu_c$ are either above or below our observing band. However, the optically thin spectral slopes at early times ($\delta t < 220$ d) are significantly steeper than those observed at later epochs (see Figure \ref{fig:spec_index}). If these early-time spectra are interpreted within the regime $\nu_m < \nu_a < \nu_c$, they would require an abnormally large electron power-law index, $p \gtrsim 3.5$.

Instead, we interpret the early-time spectra as being consistent with the regime $\nu_a > \nu_m, \nu_c$, which implies a more typical value of $p \sim 2.5$. We then attribute the flattening of the optically thin spectral slope over time to the cooling frequency, $\nu_c$, increasing from below our observed frequency range with $\nu_c<\nu_a$ to above our observed frequency range with $\nu_c>\nu_a$. The observed change in the optically thin spectral index is fully consistent with expectations for the passage of this break (see Figure \ref{fig:spec_index}). However, we note that this would imply an extraordinarily rapid evolution of $\nu_c$ across our observing band. We further discuss this assumption and its implications in Section \S \ref{sec:21sdu}. 
For now, we fit the epochs $\delta t \lesssim 220$ days assuming the spectral regime $\nu_a > \nu_m, \nu_c$, for which Equation \ref{eq:sed} gives $\beta_1 = 5/2$ and $\beta_2 = -p/2$. For the later epochs, we adopt the regime $\nu_m < \nu_a < \nu_c$, corresponding to $\beta_1 = 5/2$ and $\beta_2 = (1 - p)/2$. The spectra at later times ($\delta t > 220$ days) are then also consistent with $p\sim2.5$.

\begin{table*}[ht!]
    \centering
        \begin{tabular}{l|cccccc}
        \hline\hline
        Object & Date & $\delta t$ (days) & $F_{\nu_a,1}$ (mJy) & $\nu_{a,1}$ (GHz) & $F_{\nu_a,2}$ (mJy) &  $\nu_{a,2}$ (GHz)\\
        \hline
        \textbf{AT 2020zso} & 2021 Feb 20 &101 & $0.14\pm0.02$ & $8.61\pm2.73$ & $-$ & $-$ \\
        $p_1=2.92\pm0.08$ & 2021 May 26/28/30, Jun 08 &195-207 & $0.70\pm0.09$ & $6.00\pm1.05$ & $-$ & $-$ \\
        $p_2=3.22\pm0.07$ & 2021 Nov 2 &356 & $1.42\pm0.07$ & $3.38\pm0.20$ & $-$ & $-$ \\
        & 2022 May 12 &546 & $1.22\pm0.06$ & $1.69\pm0.10$ & $-$ & $-$ \\
        & 2022 Oct 10/Nov 2 &717-720 & $0.88\pm0.05$ & $1.13\pm0.07$ & $-$ & $-$ \\
        & 2023 Apr 13 &882 & $0.95\pm0.16$ & $0.89\pm0.21$ & $0.27\pm0.05$ & $4.95\pm0.73$ \\
        & 2023 May 27 &926 & $>0.67$ & $<0.81$ & $0.25\pm0.04$ & $3.95\pm0.42$\\
        & 2023 Oct 1 &1053& $-$ & $-$ & $0.95\pm0.21$ & $1.56\pm0.34$ \\
        & 2024 May 9 &1275& $-$ & $-$ & $0.93\pm0.08$ & $1.13\pm0.09$ \\
        & 2024 Aug 25 &1382& $-$ & $-$ & $>0.99$ & $<0.98$ \\
        \hline 
        \textbf{AT 2021sdu} & 2021 Sep 11 & 68  &  $0.30 \pm 0.01$  &	$12.65\pm0.93$ & $-$ & $-$ \\
         $p=2.63\pm0.08$ & 2021 Nov 28/Dec 21 & 146-167& $1.35\pm0.03$ & $15.89\pm0.27$ & $-$ & $-$ \\
        & 2022 Jan 26/Feb 10 & 205-220 & $1.64\pm0.03$ & $9.01\pm0.15$ & $-$ & $-$ \\
        & 2022 Apr 12/May 15 & 281-314 & $0.39\pm0.01$ & $6.24\pm0.25$ & $-$ & $-$ \\
        & 2022 Jun 19/Jul 07 & 349-367 & $0.22\pm0.02$ & $4.30\pm0.29$ & $-$ & $-$ \\
        & 2022 Sep 30/Oct 9 & 452-461 & $>0.17$ & $<1.95$ & $-$ & $-$ \\

        \hline\hline
        \end{tabular}
    \caption{Spectral energy distribution parameters from our synchrotron model. We define the location of the peak frequency and flux density ($\nu_a$, $F_{\nu_a}$) as the asymptotic intersection of the optically thick and thin power-law segments. $\delta t$ is measured with respect to the transient discovery date reported to TNS.}
    \label{tab:sed}
\end{table*} 

\subsection{MCMC Procedure}

We fit the models for AT 2020zso and AT 2021sdu outlined in Sections \S\ref{subsec:AT 2020zso_model} and \S\ref{subsec:AT 2021sdu_model} to data using the MCMC module \verb'emcee' \citep{emcee} in Python, assuming a Gaussian likelihood for the measurements. We adopt uniform priors on the model parameters $p \in [2,4]$, $\text{log}_{10}F_{\nu_a} \in [-2,2]$ and $\text{log}_{10}{\nu_a} \in [8,12]$. When fitting for the AT 2020zso epochs with a two-component SED, we imposed an additional prior $\nu_{a,1} < \nu_{a,2}$. For both TDEs, we initially fit all epochs by varying $p$, $\text{log}_{10}F_{\nu_a}$, and $\text{log}_{10}{\nu_a}$ to best determine the power law index for each shock. In AT 2020zso, we find that $p_1=2.92 \pm 0.08$ and $p_2 = 3.22 \pm 0.07$, while for AT 2021sdu, we find that $p=2.63 \pm 0.08$. We find no significant evidence for variability in $p$ within each of the three outflows, therefore we fix $p$ to these values and fit only for $F_{\nu_a}$ and $\nu_a$. We used the \verb'EnsembleSampler' from \verb'emcee' to sample the posterior distributions of the parameters $F_{\nu_a}$ and $\nu_a$ using 100 MCMC walkers. We evaluated each walker for 10,000 steps, with the first 1,000 steps discarded as burn-in. We find acceptance fractions of 20\% to 70\% for all fits, consistent with recommended efficiency rates for ensemble samplers \citep{emcee}. The best-fit SED parameters for the two components of AT 2020zso and the host-subtracted emission of AT 2021sdu are presented in Table \ref{tab:sed}, while Figure \ref{fig:dual_sed} illustrates the resulting SED fits.

\section{Outflow Model}\label{sec:analysis}
The radio emission from both AT 2020zso and AT 2021sdu reaches a peak radio luminosity of $\sim2\times 10^{38} \ \text{erg/s}$, well below the $\gtrsim 10^{40} \ \text{erg/s}$ threshold typical of relativistic on-axis jets \citep{alexander2020}. Combined with the absence of strong $\gamma$-ray or X-ray emission, this disfavors relativistic on-axis jets as the origin of the observed radio emission in both events. Instead, non-relativistic outflows offer a more plausible explanation, which is often invoked to explain the less luminous $\lesssim 10^{40} \ \text{erg/s}$ radio emission seen in many other TDEs (e.g., \citealt{alexander_2016,CNSS,at2019dsg}).

To characterize the physical properties of each outflow, we use the spectral peak of the model SEDs to estimate the size of the emitting region and place constraints on the minimum total energy of the blast wave (\citealt{1998chevalier}, \citealt{Barniol_Duran_2013}). For our analysis, we model the outflow as a non-relativistic, spherically symmetric blast wave, with the emitting region occupying a shell just behind the forward shock. We further define the location of the spectral peak as the asymptotic intersection of the optically thick and thin power-law segments.

\begin{figure*}[t!]
    \centering
    \includegraphics[width = \textwidth]{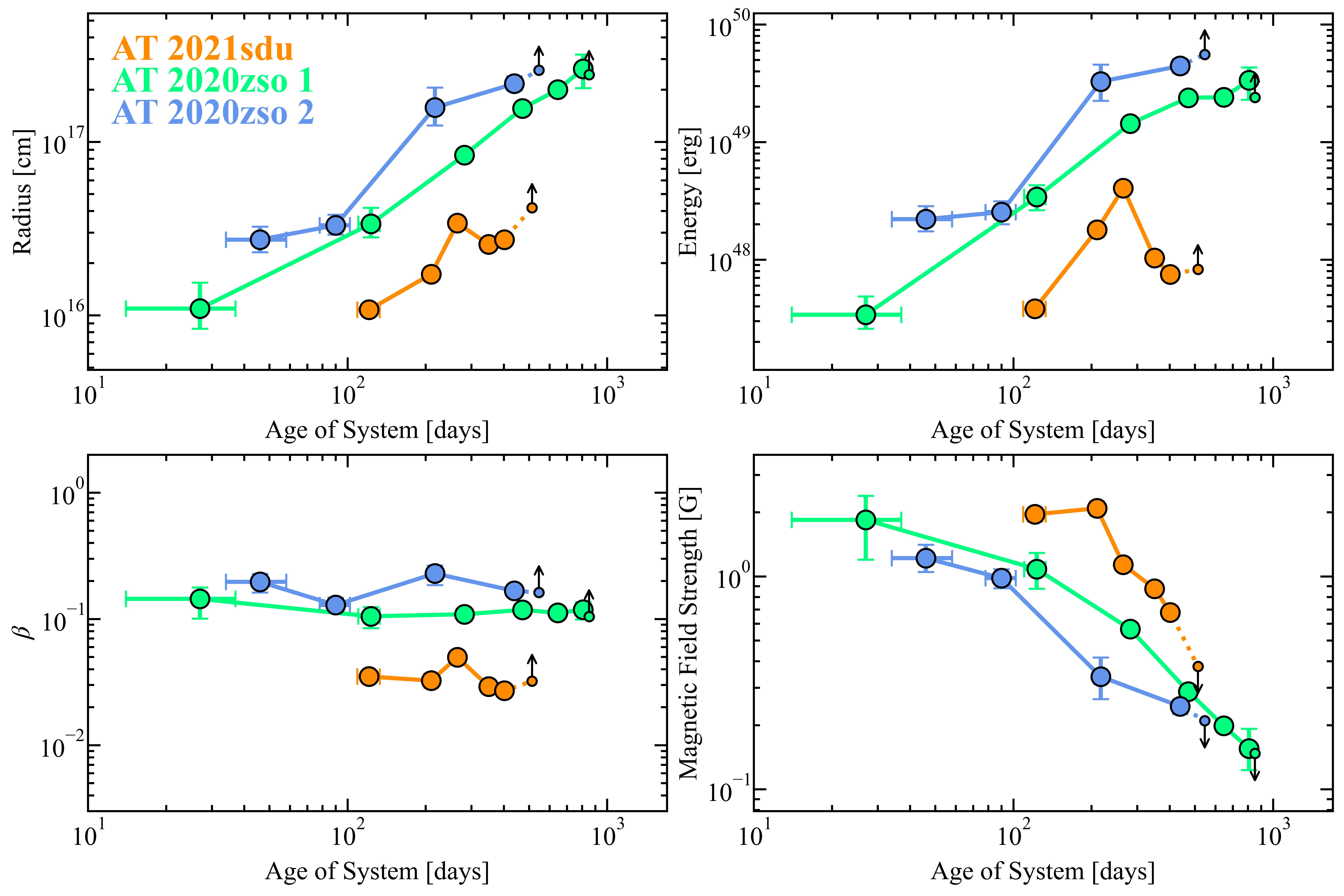}
    \caption{Evolution of various physical parameters as a function of the system's dynamical age, $\delta t-t_{0,R}$. \textit{(Top, Left)} Radius, \textit{(Top, Right)} Energy, \textit{(Bottom, Left)} Velocity, and \textit{(Bottom, Right)} Magnetic Field Strength for AT 2020zso's first (green) and second (blue) outflow, as well as AT 2021sdu's outflow (orange). Solid markers represent measured parameters, whereas open markers indicate limits when the spectral peak of the synchrotron emission is not captured.}

    \label{fig:eq_plots}
\end{figure*}

Table \ref{tab:eq} summarizes the derived outflow parameters, including radius $R$, energy $E$, velocity $\beta$, number of emitting electrons $N_e$, magnetic field strength $B$, and the CNM electron density $n_{\text{ext}}$. Additionally, we display these results in Figure \ref{fig:eq_plots}. These quantities are computed following the scaling relations from \citet{Barniol_Duran_2013}, given below.

The peak synchrotron frequency $\nu_p$ and associated flux density $F_{p}$ are related to the emitting region size $R$ and total energy $E$ by:

\begin{align}
\label{eq:R}
    R \approx& \ 1\times10^{17} \text{cm} \left(21.8(525)^{p-1}\right)^{\frac{1}{13+2p}} \nonumber \\
      & \times \gamma_m^{\frac{2-p}{13+2p}} \left(\frac{F_p}{\text{mJy}}\right)^{\frac{6+p}{13+2p}} \left(\frac{d_L}{10^{28}\text{cm}}\right)^{\frac{2(6+p)}{13+2p}} \nonumber \\
      & \times  \left(\frac{\nu_p}{10 \text{ GHz}}\right)^{-1} (1+z)^{-\frac{19+3p}{13+2p}}  \nonumber\\
      & \times  f_A^{-\frac{5+p}{13+2p}} f_V^{-\frac{1}{13+2p}} 4^{\frac{1}{13+2p}} \xi^{\frac{1}{13+2p}} \epsilon^\frac{1}{17}
\end{align}

\begin{align}
\label{eq:E}
    E \approx& \ 1.3\times10^{48} \text{erg}\left(21.8\right)^{-\frac{2(p+1)}{13+2p}}\left(525^{p-1}\gamma_m^{2-p}\right)^{\frac{1}{13+2p}} \nonumber \\
      & \times \left(\frac{F_p}{\text{mJy}}\right)^{\frac{14+3p}{13+2p}} \left(\frac{d_L}{10^{28}\text{cm}}\right)^{\frac{2(3p+14)}{13+2p}} \left(\frac{\nu_p}{10 \text{ GHz}}\right)^{-1} \nonumber \\
      & \times (1+z)^{-\frac{27+5p}{13+2p}} f_A^{-\frac{3(p+1)}{13+2p}} f_V^{\frac{2(p+1)}{13+2p}} \nonumber  \\
      & \times 4^{\frac{1}{13+2p}} \xi^{\frac{1}{13+2p}}(\frac{11}{17}\epsilon^\frac{-6}{17} +\frac{6}{17}\epsilon^\frac{11}{17})
\end{align}

\noindent where $d_L$ is the luminosity distance for a source at redshift $z$; $f_A=1$ and $f_V\approx0.36$ are the area and volume filling factors where we take the emitting region to be a 0.1 $R$ thick spherical shell just behind the shock front (e.g., \citealt{alexander_2016}; \citealt{at2019dsg}; \citealt{Christy_2024}); $\xi = 1 + \epsilon_e^{-1}$ describes the contribution of hot protons; and $\epsilon=(11/6)(\epsilon_B/\epsilon_e)$ describes potential deviations from equipartition where $\epsilon_e$, $\epsilon_B$ are the fractions of post-shock energy density in electrons and magnetic fields, respectively. For our analysis, we assume equipartition where $\epsilon_e=\epsilon_B=0.1$. Additionally, we set $\gamma_m = 2$ and include the relevant factors of 4 to correct the isotropic number of radiating electrons ($N_e,\text{iso}$,iso) in the Newtonian limit.

Given an estimate of $R$, we can derive the Lorentz factor of the electrons emitting near peak, $\gamma_e$ by,
\begin{align}
\label{eq:g}
    \gamma_e \approx& \ 525 \left(\frac{F_p}{\text{mJy}}\right) \left(\frac{d_L}{10^{28}\text{cm}}\right)^{2}  \left(\frac{\nu_p}{10 \text{ GHz}}\right)^{-2} \nonumber \\
    & \times (1+z)^{-3} f_A^{-1} \left(\frac{R}{10^{17} \text{ cm}}\right)^{-2}
\end{align}

\noindent The magnetic field strength and number of electrons contained within the emitting region can be expressed as follows,

\begin{align}
\label{eq:B}
    B \approx& \ 1.3\times10^{-2} \text{G} \left(\frac{F_p}{\text{mJy}}\right)^{-2} \left(\frac{d_L}{10^{28}\text{cm}}\right)^{-4} \nonumber \\
    & \times \left(\frac{\nu_p}{10 \text{ GHz}}\right)^{5} (1+z)^{7} f_A^{2} \left(\frac{R}{10^{17} \text{ cm}}\right)^{4}
\end{align}

\begin{align}
\label{eq:N}
    N_e \approx& \ 4\times10^{54} \left(\frac{F_p}{\text{mJy}}\right)^{3} \left(\frac{d_L}{10^{28}\text{cm}}\right)^{6} \left(\frac{\nu_p}{10 \text{ GHz}}\right)^{-5} \nonumber \\
    & \times (1+z)^{-8} f_A^{-2} \left(\frac{R}{10^{17} \text{ cm}}\right)^{-4} \left(\frac{\gamma_e}{\gamma_m}\right)^{p-1} 
\end{align}

\noindent Since $\nu_m < \nu_a$, a significant portion of the electrons' energy is carried by electrons whose emission is self-absorbed. We account for this by including correction factors of $\left(\gamma_m/\gamma_e\right)$ to $N_e$ and subsequently the total electron energy. These corrections are already included in the expressions for the radius and energy presented above. 

Given $N_e$ and $R$, we can estimate the CNM as a function of radius. We define the ambient electron density at radius $R$ as $n_{\text{ext}}=n_e/4$, where $n_e=N_e/V$ is the number density of electrons in the emitting region of the outflow, with an emitting volume $V=f_V\pi R^3$. The factor of 4 in the density estimate accounts for the jump conditions across the shock front for adiabatic index $\gamma=5/3$.

\begin{figure}[t!]
    \centering
    \includegraphics[width = 0.5\textwidth]{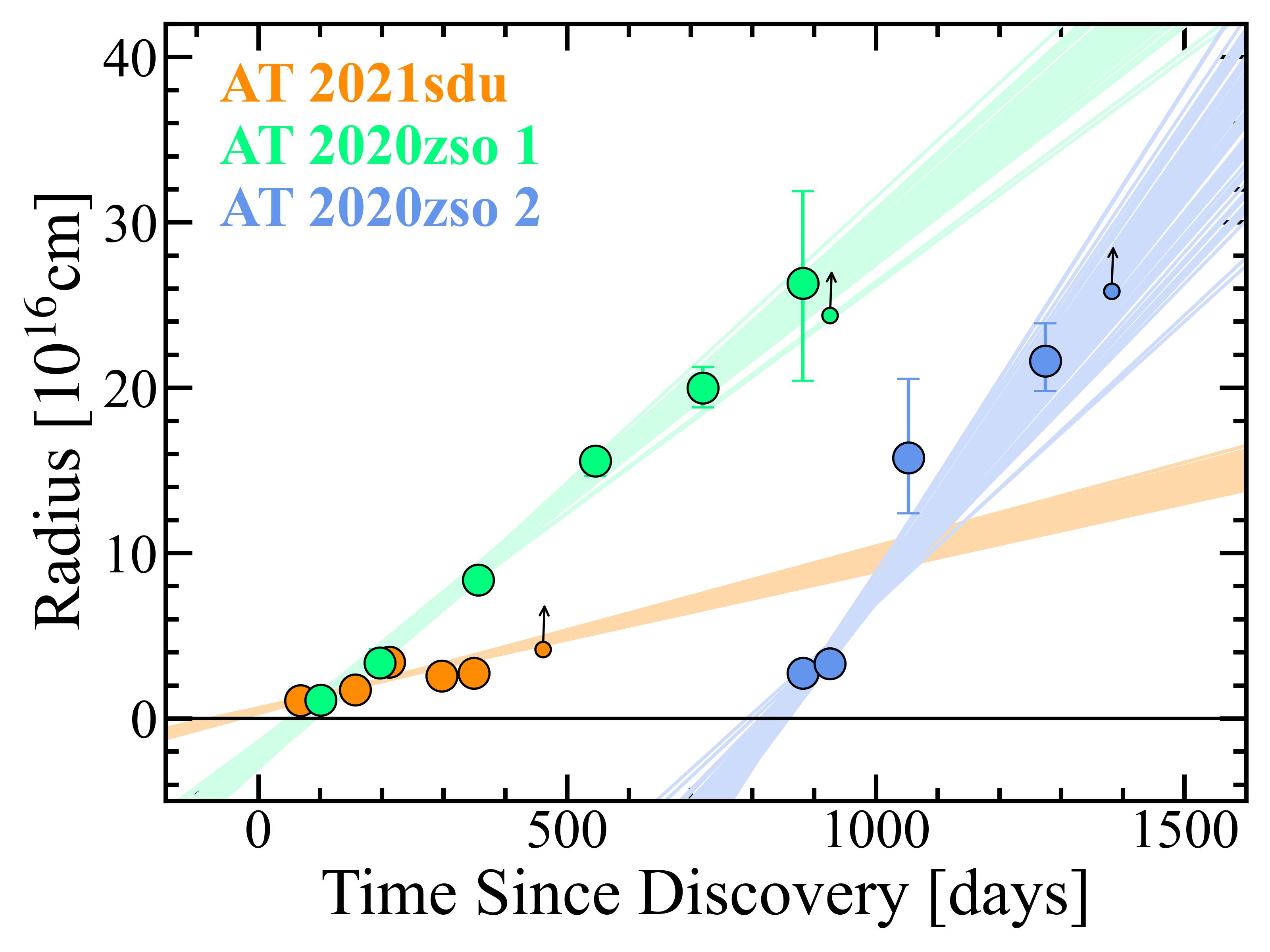}
    \caption{Evolution of the emitting region size for AT 2020zso's first (green) and second (blue) outflow, and AT 2021sdu's outflow (orange) with respect to the transient discovery date. Points with arrows represent inferred lower limits. Solid lines represent linear fits to the data (the expectation for a free expansion blast wave). By extrapolating the linear fits, we find that the following radio outflow launch dates: $t_{0,R}=74_{-12}^{+12}$ d for the first component in AT 2020zso; $t_{0,R}=836_{-13}^{+10}$ d for the second component in AT 2020zso; and $t_{0,R}=-53_{-12}^{+12}$ d for AT 2021sdu.}
    \label{fig:launch}
\end{figure}

With enough temporal coverage, we can place tight constraints on the launch date of the outflow by linearly extrapolating $R(t)\xrightarrow{}0$. This procedure is valid under the assumption that the blast wave has remained in free expansion. In Figure \ref{fig:launch}, we show linear fits to $R(t)$ for AT 2021sdu and the first and second outflows in AT 2020zso. The time since the outflow launch date can be expressed as $\delta t-t_{0,R}$, where $\delta t$ is the time elapsed between the transient's optical discovery date and the observation; while $t_{0,R}$ is the launch date inferred from the radius evolution (measured relative to the optical discovery date). With this, we can infer the free expansion velocity by relating $t=\delta t-t_{0,R}$ to the radius of the blast wave as follows:
\begin{equation}
    t = \frac{R(1+z)(1-\beta)}{c \beta}
\end{equation}

\noindent where $\beta$ is the source velocity assuming no deceleration \citep{Barniol_Duran_2013}.

\begin{table*}
    \centering
    \begin{tabular}{l|cccccccc}
        \hline\hline
        Object & $\delta t$ & $\delta t - t_{0,R}$ & $\log_{10}R$ & $\log_{10}E$ & $\log_{10}B$ & $\log_{10}N_e$ & $\log_{10}n_\text{ext}$ & $\beta$ \\
        
         & (d) &(d) & (cm) & (erg) & (G) &  & (cm$^{-3}$) & \\
        \hline
         \textbf{AT 2020zso} & 101 & 27$_{-12}^{+12}$& $16.04_{-0.10}^{+0.18}$ & $47.53_{-0.10}^{+0.19}$ & $0.27_{-0.15}^{+0.13}$ & $51.90_{-0.10}^{+0.19}$ & $3.13_{-0.26}^{+0.30}$ & $0.135_{-0.032}^{+0.038}$ \\
         $t_{0,R}=74_{-12}^{+12}$ d & 197 & 123$_{-12}^{+12}$& $16.53_{-0.07}^{+0.10}$ & $48.53_{-0.10}^{+0.11}$ & $0.03_{-0.08}^{+0.08}$ & $52.90_{-0.10}^{+0.11}$ & $2.66_{-0.16}^{+0.19}$ & $0.101_{-0.014}^{+0.018}$ \\
         & 356 & 282$_{-12}^{+12}$& $16.92_{-0.03}^{+0.03}$ & $49.16_{-0.03}^{+0.04}$ & $-0.25_{-0.02}^{+0.03}$ & $53.53_{-0.03}^{+0.04}$ & $2.10_{-0.05}^{+0.05}$ & $0.109_{-0.006}^{+0.006}$ \\
         & 546 & 472$_{-12}^{+12}$& $17.19_{-0.03}^{+0.02}$ & $49.38_{-0.03}^{+0.03}$ & $-0.54_{-0.02}^{+0.02}$ & $53.75_{-0.03}^{+0.03}$ & $1.51_{-0.04}^{+0.05}$ & $0.119_{-0.006}^{+0.006}$ \\
         & 720 & 646$_{-12}^{+12}$& $17.3_{-0.03}^{+0.03}$ & $49.38_{-0.04}^{+0.04}$ & $-0.70_{-0.03}^{+0.02}$ & $53.75_{-0.04}^{+0.04}$ & $1.19_{-0.05}^{+0.03}$ & $0.113_{-0.006}^{+0.006}$\\
         & 882 & 808$_{-12}^{+12}$& $17.42_{-0.10}^{+0.09}$ & $49.53_{-0.14}^{+0.12}$ & $-0.81_{-0.09}^{+0.10}$ & $53.90_{-0.14}^{+0.12}$ & $0.98_{-0.17}^{+0.24}$ & $0.113_{-0.017}^{+0.02}$\\
         & 926 & 852$_{-12}^{+12}$& $>17.39$ & $>49.38$ & $<-0.83$ & $>53.75$ & $<0.93$ & $>0.104$ \\
        \hline
         \textbf{AT 2020zso} & 882 & 46$_{-13}^{+10}$ & $16.44_{-0.07}^{+0.08}$ & $48.34_{-0.09}^{+0.13}$ & $0.09_{-0.06}^{+0.07}$ & $52.69_{-0.09}^{+0.13}$ & $2.73_{-0.11}^{+0.15}$ & $0.195_{-0.026}^{+0.028}$ \\
         $t_{0,R}=836_{-13}^{+10}$ d& 926 & 90$_{-13}^{+10}$ & $16.52_{-0.05}^{+0.07}$ & $48.40_{-0.09}^{+0.10}$ & $-0.01_{-0.04}^{+0.04}$ & $52.75_{-0.09}^{+0.10}$ & $2.54_{-0.08}^{+0.09}$ & $0.128_{-0.012}^{+0.013}$ \\
         & 1053 & 217$_{-13}^{+10}$ & $17.20_{-0.09}^{+0.13}$ & $49.51_{-0.14}^{+0.17}$ & $-0.47_{-0.09}^{+0.10}$ & $53.86_{-0.14}^{+0.17}$ & $1.61_{-0.17}^{+0.22}$ & $0.229_{-0.042}^{+0.043}$\\
         & 1275 & 439$_{-13}^{+10}$ & $17.33_{-0.04}^{+0.05}$ & $49.65_{-0.06}^{+0.06}$ & $-0.61_{-0.03}^{+0.03}$ & $54.00_{-0.06}^{+0.06}$ & $1.33_{-0.07}^{+0.07}$ & $0.167_{-0.011}^{+0.013}$\\
         & 1382 & 546$_{-13}^{+10}$ & $>17.41$ & $>49.75$ & $<-0.68$ & $>54.09$ & $<1.20$ & $>0.162$ \\
        \hline
        
        \textbf{AT 2021sdu} & 68 & 121$_{-12}^{+12}$ & $16.03_{-0.03}^{+0.04}$ & $47.58_{-0.04}^{+0.04}$ & $0.29_{-0.03}^{+0.03}$ & $51.97_{-0.04}^{+0.04}$ & $3.22_{-0.06}^{+0.07}$ & $0.035_{-0.003}^{+0.002}$ \\
         $t_{0,R}=-53_{-12}^{+12}$ d& 157 & 210$_{-12}^{+12}$ & $16.24_{-0.01}^{+0.01}$ & $48.25_{-0.01}^{+0.01}$ & $0.32_{-0.01}^{+0.01}$ & $52.65_{-0.01}^{+0.01}$ & $3.28_{-0.01}^{+0.02}$ & $0.033_{-0.001}^{+0.001}$\\
         & 212 & 265$_{-12}^{+12}$ & $16.53_{-0.01}^{+0.01}$ & $48.61_{-0.01}^{+0.01}$ & $0.05_{-0.01}^{+0.01}$ & $53.00_{-0.01}^{+0.01}$ & $2.75_{-0.01}^{+0.01}$ & $0.050_{-0.001}^{+0.001}$ \\
         & 297 & 350$_{-12}^{+12}$ & $16.41_{-0.02}^{+0.02}$ & $48.01_{-0.02}^{+0.02}$ & $-0.06_{-0.02}^{+0.02}$ & $52.41_{-0.02}^{+0.02}$ & $2.52_{-0.03}^{+0.03}$ & $0.029_{-0.001}^{+0.001}$ \\
         & 349 & 402$_{-12}^{+12}$ & $16.44_{-0.03}^{+0.03}$ & $47.87_{-0.04}^{+0.05}$ & $-0.17_{-0.03}^{+0.03}$ & $52.27_{-0.04}^{+0.05}$ & $2.30_{-0.06}^{+0.06}$ & $0.027_{-0.002}^{+0.002}$\\
         & 461 & 514$_{-12}^{+12}$ & $>16.62$ & $>47.92$ & $<-0.42$ & $>52.31$ & $<1.79$ & $>0.032$ \\

        \hline\hline

    \end{tabular}
    \caption{Results of our equipartition analysis described in Section \S \ref{sec:analysis}. When computing $\beta$, we use $\delta t-t_{0,R}$ as the dynamical age of the system, where $t_{0,R}$ is the inferred launch date of the outflow measured with respect to the optical discovery date.}
    \label{tab:eq}  
\end{table*}

\section{Discussion}\label{sec:disc}
In this section, we discuss the possible outflow launching mechanisms responsible for the radio emission in AT 2020zso and AT 2021sdu. We also examine the physical evolution of the outflows as inferred from our equipartition analysis.

\subsection{AT 2020zso: The first outflow}
The evolution of the emission radius for the first flare in AT 2020zso implies a launch date of $t_{0,R_1} = 74_{-12}^{+12}$ days relative to optical discovery. This corresponds to $\sim56$ days after the peak of the bolometric light curve reported in \citet{Wevers_2022}. We also find that this outflow is non-relativistic with $v\approx0.11c$ and a total internal energy plateauing near $E\sim2.2\times10^{49}$ erg (see Figure \ref{fig:eq_plots}), consistent with other non-jetted TDEs \citep{alexander_2016,at2019dsg,AT2020vwl}.

The earliest outflows in TDEs may be launched shortly after the disruption itself, when roughly half of the stellar material becomes unbound from the SMBH. As the fastest-moving portion of the unbound debris interacts with the CNM, a bow shock forms at the leading edge of the debris stream, which can generate synchrotron emission \citep{krolik_2016}. The maximum velocity of the unbound debris depends on the impact parameter $\beta$, defined as the ratio of the tidal radius to the pericenter distance. For deeply penetrating encounters (e.g., $\beta=7$), the fastest debris is expected to move at $v\lesssim0.06c$ \citep{Yalinewich_2019}, while more shallow encounters ($\beta\sim1$) produce even slower ejecta. Modeling of the optical emission in AT~2020zso suggests an impact parameter of order unity, consistent with a shallow encounter \citep{Wevers_2022}.

In contrast, the outflow velocity we infer for the first flare in AT 2020zso ($v\approx0.11c$) is higher than predicted in most unbound debris models \citep{krolik_2016,Yalinewich_2019}, and the inferred energy of the outflow ($E\sim2.2\times10^{49}$ erg) also exceeds the $\sim10^{47}-10^{48}$ erg proposed for the unbound debris stream \citep{krolik_2016}. Adding to this, we find that the blast wave is launched months after the optical transient first appears, whereas unbound debris would be expected to be launched around the time of the star's initial disruption. Therefore, we conclude that unbound debris is unlikely to be the origin of this outflow.

After the unbinding of the star, outflows can also form during the circularization of the bound debris. \citet{Lu_2020} showed that shocks at the self-intersection of the bound debris streams can unbind and eject a substantial amount of gas (referred to as collisionally-induced outflows, i.e., CIOs). The energy and velocity of the prompt outflow in AT 2020zso are consistent with the theoretical expectation of CIOs, which would have total kinetic energies of $\sim10^{50}$ erg and velocities between $0.01c - 0.1c$. The CIO scenario proposes that the optical emission in TDEs is the result of reprocessing of the higher-energy emission from accretion via outflows generated during the stream-stream interactions \citep{Lu_2020}. Under the assumption that the optical emission is entirely driven by stream-stream collisions, any associated radio-emitting outflows should be produced at a similar time. However, our inferred outflow launch date occurs months after the onset of the optical flare and weeks after peak optical light. While this temporal offset challenges a direct stream-stream collision origin, uncertainties in the outflow launch epoch prevent us from conclusively ruling out this scenario. 

The radio emission could also arise from an accretion-driven process associated with the fallback of the bound debris, which is expected to be initially super-Eddington \citep{Rees1998}. During this phase,  radiation pressure is expected to drive winds from the accretion disk \citep{Strubbe}. These winds may shock the surrounding CNM, generating observable radio emission \citep{ Mohan_2022}. Such disk winds represent a plausible mechanism for producing radio emission in TDEs lacking evidence for a relativistic jet.

Evidence for this scenario is present in AT 2020zso. Near optical peak (from $\delta t\approx 15-28$ d), the optical spectra displayed H$\alpha$ and  He~\rom{2} line profiles indicative of a highly inclined compact accretion disk \citep{Wevers_2022}. This implies that accretion onto the SMBH had already begun by this time, disfavoring the CIO scenario where the peak optical emission is dominated by stream-stream collisions. Our analysis indicates the first radio flare was launched shortly after this time. Additionally, the peak UV/optical luminosity corresponded to an Eddington ratio of $\sim$0.5, which suggests that the accretion disk was geometrically slim \citep{Abramowicz_1988,Wevers_2022}.  Simulations of accretion disks with this geometry indicate that they are  capable of driving powerful outflows (e.g., \citealt{Sadowski_2015}). Therefore, it is plausible that the first radio flare was accretion-driven in origin due to this temporal coincidence. Moreover, the outflow expansion velocity of $v\approx0.11c$ and energy of $E\sim2.2\times10^{49}$ erg are also consistent with those expected from an accretion-driven wind (e.g., \citealt{Strubbe}). Although no transient X-ray emission was detected \citep{Wevers_2022}, which would typically signal ongoing accretion onto the black hole, such non-detections are consistent with an edge-on viewing geometry, which can obscure inner disk emission due to self-obscuration or absorption by a geometrically thick, optically dense accretion flow \citep{Dai_2018}.

In summary, we favor an accretion-driven outflow as the origin of the first radio flare. While the dynamical evolution is broadly consistent with either an accretion-driven or CIO scenario, the presence of a newly formed TDE disk near the time of the outflow launch date strongly supports the accretion-driven picture. By contrast, CIOs are expected much earlier in the disruption process as they rely on collisions between the bound debris streams prior to circularization. In a scenario where a disk has already been formed, it is unlikely that CIOs would arise months after the fact.

\subsection{AT 2020zso: The second outflow}
Our light curve and spectral modeling both indicate the presence of an emerging secondary outflow which begins to dominate the radio spectra around $\delta t\sim900$ days. By extrapolating the emitting region size, we find that this outflow is consistent with a launch date of $t_{0,2} = 836_{-13}^{+10}$ days. Similar to the first radio flare, the late-time flare exhibits properties consistent with those of a freely coasting blast wave where $v\approx0.16c$. The total internal energy of this outflow component reaches $E\approx3.8\times10^{49} \text{ erg}$, comparable to the energetics of the first outflow. This puts AT 2020zso in the class of TDEs that exhibit both an early-time ($\delta t < 1 \text{ yr}$) radio flare and a re-brightening at late-times ($\delta t > 1 \text{ yr}$). Distinct early and late-time radio emission has been observed in only a few TDEs: ASASSN-15oi (\citealt{horesh_15oi,Hajela_2024}), AT 2019azh \citep{sfaradi_2022}, AT 2019dsg \citep{Cendes_2024}, ASASSN-19bt \citep{Christy_2024}, and AT 2020vwl \citep{Goodwin_2025}. 

Since we resolve two distinct components in the radio spectra of AT 2020zso, we are likely seeing the contribution of two separate emitting regions. This aspect of our radio data rules out energy injection into the first outflow as the origin of the second flare (one of the scenarios proposed for AT 2020vwl in \citealt{Goodwin_2025}), and instead points to the second outflow as distinct from the first. This leaves us with a few possible scenarios that we discuss below.

\subsubsection{Decelerating Off-Axis Relativistic Jet}
This second radio flare emerges on a timescale similar to the delayed outflow seen in AT 2018hyz. The onset of this flare was detected around 972 days after the optical discovery (\citealt{AT2018hyz}). One possible model previously proposed for AT 2018hyz is that the delayed flare arises from a promptly launched off-axis relativistic jet \citep{Sfaradi_2024,matsumoto}. In this case, the observed radio emission only becomes visible once the relativistic beaming cone has widened sufficiently, due to jet deceleration, allowing the emission to reach our line of sight (\citealt{Sfaradi_2024, matsumoto}).

A key parameter for determining if this model is correct is the value and evolution of the apparent outflow velocity $\beta_\text{app}$. For superluminal apparent velocities ($\beta_\text{app}>1$), the emitting source is relativistic. However, \citet{matsumoto} showed that apparent velocities $\beta_{\rm{app}}\ll1$ are not necessarily indicative of sources that are Newtonian. Below a critical apparent velocity $\beta_{\mathrm{crit}} < 0.44$, the possible solutions for the true velocity of the source separate into relativistic and Newtonian regimes \citep{Beniamini_2023}. We find that this second radio flare in AT 2020zso exhibits an apparent expansion velocity of $\beta_\text{app}\approx0.16$, which is below this critical threshold, so an off-axis jet interpretation may be feasible. If the off-axis interpretation holds, $\beta_{\mathrm{app}}$ should rise above the critical value when the emission is off-axis and then decline once fully on-axis. However, we observe that the apparent velocity instead remains constant, inconsistent with the predicted evolution of an emerging off-axis jet. This evolution more closely aligns with that of a non-relativistic blast wave. Moreover, the light curves for the delayed flare in AT 2020zso and AT 2018hyz differ in their evolution. AT 2018hyz displayed a rapid and long-lasting rise with $F_\nu\propto t^5$, whereas the second flare in AT 2020zso has already begun fading for frequencies $\nu\gtrsim3 \text{ GHz}$. Therefore, we disfavor a delayed off-axis relativistic jet as the origin of the second radio flare.

\subsubsection{Delayed Accretion}
A delayed radio flare in TDEs may signal late-time accretion activity, a process hinted at by multi-wavelength observations in other events (e.g., \citealt{liu_2022,Hajela_2024,Alexander_2025}). If the prompt radio emission originates from CIOs, a second flare could arise from delayed accretion, a scenario supported by systems like ASASSN-15oi, AT 2019azh, and AT 2020vwl. In ASASSN-15oi, \citet{Hajela_2024} linked the delayed radio flare ($\gtrsim200$ days post-optical peak) to concurrent X-ray detections, suggesting accretion-driven processes (e.g., disk winds) during peak accretion. Similarly, AT 2019azh exhibited a late-time X-ray flare $\sim200$ days post-discovery, coinciding with renewed radio activity \citep{liu_2022}. For AT 2020vwl, \citet{Goodwin_2025} modeled the accretion rate over time and similarly found that the time of peak accretion was consistent with the date of the second radio flare ($\sim160 - 700$ days post-discovery). 

These cases establish a framework where delayed radio emission is consistent with a delayed high accretion state. For AT 2020zso, the second flare's timing ($t_{0,R_2} \sim 836$ d) superficially mirrors this pattern. Unlike other systems that show both delayed radio flares and signs of delayed accretion, AT 2020zso has no multi-wavelength counterpart at late times. As previously discussed, AT 2020zso was never detected in the X-rays. Similarly, no plateau phase was seen in UV/optical observations (\citealt{Wevers_2022,Mummery_2024,Franz_2025}). Due to this paucity of data, it is difficult to probe the accretion rate around the time of the second radio flare. However, if AT 2020zso already hosted a compact accretion disk near optical peak (as suggested by \citealt{Wevers_2022}), it would be difficult to reconcile this scenario with interpreting the second radio flare as the result of delayed accretion.

\subsubsection{Prompt Accretion Followed by a Late-Time State Transition}

Alternatively, a delayed radio flare may not indicate delayed peak accretion. Instead, the fallback rate onto the SMBH could have increased rapidly and been initially super-Eddington, as expected in many TDEs~\citep{Rees1998}, and then declined over time. During this early super-Eddington phase, the accretion disk can launch non-relativistic outflows \citep{Strubbe}, potentially producing the prompt radio emission we observe in AT 2020zso. Once the accretion rate drops below  $\dot{M} \lesssim0.3\dot{M}_{\text{Edd}}$, the disk transitions to a geometrically thin state, which is not expected to produce outflows \citep{Tchekhovskoy}. At later times, when the fallback rate falls further to $\dot{M}\lesssim0.03\dot{M}_{\text{Edd}}$), the accretion flow may transition to a geometrically thick, Advection-Dominated Accretion Flow (ADAF) regime \citep{van_Velzen_2011,Giannios_2011}. In the context of X-ray binaries, this is referred to as the ``low/hard'' state, which is linked to the launching of outflows \citep{Fender_2004}.

This evolution motivates a scenario in which the first radio flare is powered by a super-Eddington disk wind (or potentially a stream–stream CIO if the fallback rate never became super-Eddington), while the second, delayed radio flare arises from a state change to an analogous low-hard state. In this picture, the delayed flare reflects not renewed accretion, but rather a change in the accretion flow and its ability to launch an outflow as the overall accretion rate continues to fall.

\subsubsection{Partial Disruption Scenario}
The above scenarios all interpret the two radio outflows within the context of a single TDE. However, light curve modeling of the UV/optical emission suggests that AT 2020zso may be a partial TDE, inferred from its low impact parameter and minimal debris mass \citep{Wevers_2022}. Other candidate partial TDEs have been shown to repeat, where a second optical flare repeats during the second flyby of the partially disrupted star \citep{Somalwar_2023, Lin_2024}. The timescale corresponding to the delay between the first and second optical flare in these events is comparable to the time between the first and second radio flares in AT 2020zso. Unfortunately, AT 2020zso was sun-constrained during the estimated second radio flare launch date ($\delta t \approx 836~\rm{days}$   on 2023 Feb 26). Therefore, it is unlikely that a corresponding optical flare would have been captured.

If AT 2020zso is in fact a partial TDE, the similarity in energy and velocity between the two radio outflows ($E \sim 10^{49}$ erg, $v \sim 0.11c$–$0.16c$) may imply a shared physical origin. This would suggest that periodic outflows on a $\sim800$-day timescale would continue until the surviving stellar core becomes fully torn apart by the SMBH. However, due to the lack of any coincident optical flare, we remain agnostic to this interpretation. Future observations at radio and optical wavelengths will test this scenario.

In summary, we find that AT 2020zso's first radio flare is best explained by an accretion-driven outflow while the TDE disk was accreting at a relatively high Eddington fraction. Evidence for prompt accretion in AT 2020zso is also supported by the findings of \citet{Wevers_2022}. These results, combined with the lack of renewed optical activity during the second radio flare, suggest that the delayed flare is best explained not by a second episode of rapid accretion, but by a state transition in the accretion flow that re-enabled jet or wind launching as the accretion rate declined to a highly sub-Eddington regime.

\subsection{AT 2021sdu: Ejection of Unbound Debris?}\label{sec:21sdu}

We detected both early- and late-time radio emission from AT 2021sdu. However, rather than two distinct transient components, our observations suggest the presence of a single radio transient superimposed on a persistent, non-variable host component. The presence of this host emission limits our ability to monitor the transient, which becomes indistinguishable from the host around $\delta t \sim500\text{ d}$. Prior to this, we find that the transient emission can be well modeled as an outflow that is overall less energetic and slower than those powering the radio emission from AT 2020zso. Our equipartition analysis suggests a total kinetic energy of $E\sim10^{48}\text{ erg}$ and an outflow velocity of $v\approx0.03c$. Interestingly, we find that the radial evolution of the blast wave implies a launch date of $t_{0,R}\approx-53_{-12}^{+12}\text{ d}$ with respect to the transient discovery date (see Figure \ref{fig:launch}). The only mechanism that could launch an outflow this early on in the disruption process without generating a coincident optical flare would be the ejection of the unbound stellar debris. However, the inferred early launch date could also result from a non-constant outflow velocity or an interaction with a dense CNM.

When the data are considered in more detail, however, a simple outflow picture begins to break down around $\delta t \gtrsim 220$ days. In particular, we observe a factor of $\sim$4 decrease in the peak flux density between $\delta t \approx 220$ and 314 days, which yields an unphysical blast wave solution, namely a reversal in the outflow radius. This is unlikely to be impacted by host contamination as the peak flux density is orders of magnitude larger than the corresponding host flux density at the same frequency. A similar phenomenon was observed in AT 2020vwl, where the equipartition radius decreased briefly. This was interpreted as evidence for variable energy injection, or different emitting regions dominating at different times, as an asymmetric outflow encountered a clumpy CNM \citep{AT2020vwl}. A similar interpretation may be viable here.

Another possible explanation is that one or more assumptions underlying our equipartition analysis are invalid, particularly the assumption that the emission geometry stays constant throughout our observations. The geometry of the emitting region, with area $A$ and volume $V$, is parameterized by the filling factors $f_A=A/(\pi R^2/\Gamma^2)$ and $f_V=V/(\pi R^3/\Gamma^4)$. These factors can significantly impact the inferred radius, since $R \propto f_A^{-0.42} f_V^{-0.05}$ for $p = 2.63$. We adopt $f_A = 1$ and $f_V = 0.36$, which correspond to a uniformly emitting shell just behind the forward shock. However, if the emission becomes patchier at late times (e.g., due to turbulence or fragmentation), a modest reduction in $f_A$ and $f_V$ could restore the expected monotonic radius evolution.

Another assumption baked into our estimation of the shock radius was that $\epsilon_e=\epsilon_b$ and both remain constant. We find that varying these microphysical parameters does not alleviate the discrepancy. To recover a monotonically increasing radius, $\epsilon_e$ would need to decrease by a factor of $\sim100$, or $\epsilon_B$ would have to become greater than one, both of which are physically implausible.

Our early SEDs require $\epsilon_B\gtrsim 0.01$ to match $\nu_c < \nu_a \lesssim10~\text{GHz}$ (Section \ref{subsec:AT 2021sdu_model}), consistent with our equipartition assumption. However, as with our radius estimates, this agreement breaks down at later times. By the later epochs where $\delta t \gtrsim 314$ days, we observe a shift in the optically thin spectral index consistent with $\nu_c$ moving from $\nu_c < \nu_a$ to above our observing bands (Section \ref{subsec:AT 2021sdu_model}). Although our early-time estimate of $\epsilon_B$ aligns with $\nu_c$, it cannot account for the rapid evolution of $\nu_c$ seen in the data ($\nu_c \lesssim10~\text{GHz}$ to $\nu_c \gtrsim25~\text{GHz}$ over $\sim100$ days). Exploring alternative values for $\epsilon_B$ does not resolve this behavior, indicating that, similar to the radius evolution issue, there are limitations to the simplified assumptions in our equipartition analysis. A more complete treatment of the shock microphysics and emission geometry may therefore be required to fully account for the observed spectral evolution, but this added complexity would require fine-tuning of additional free parameters that are degenerate with respect to the current data and we therefore do not pursue it here.

The signs of variable geometrical factors may point to changes in the outflow structure itself. One possibility is gravitational fragmentation of the unbound debris, as predicted in some simulations \citep[e.g.,][]{Coughlin_2016}. In this picture, the observed late-time evolution naturally results from the fragmentation of an unbound debris-driven outflow. However, others have proposed that the radio emission is driven by only the fastest-moving portion of the unbound debris (e.g., \citealt{Yalinewich_2019}). In this scenario, the debris responsible for the emission is not expected to clump or collapse, which weakens the interpretation of a changing source geometry in an unbound debris-driven outflow. 

If we allow a changing source structure where $f_A$ and $f_V$ decrease at late times, the inferred launch date becomes closer to coinciding with the time of the optical flare, making a CIO and/or disk wind origin more plausible. In such scenarios, one might also expect strong X-ray emission at early times as a signature of active accretion. However, \citet{Guolo_24b} was unable to place meaningful constraints due to the high Galactic absorption toward AT 2021sdu ($N_{\rm H,G} \geq 10^{21} \ \text{cm}^{-2}$), making it difficult to directly probe accretion activity. We therefore conclude that the outflow may plausibly originate from either the unbound debris or from stream collisions/disk winds, provided that the emitting region becomes increasingly inhomogeneous at late times.

\subsection{Properties of the Circumnuclear Medium}

\begin{figure}[t!]
    \centering
    \includegraphics[width = 0.5\textwidth]{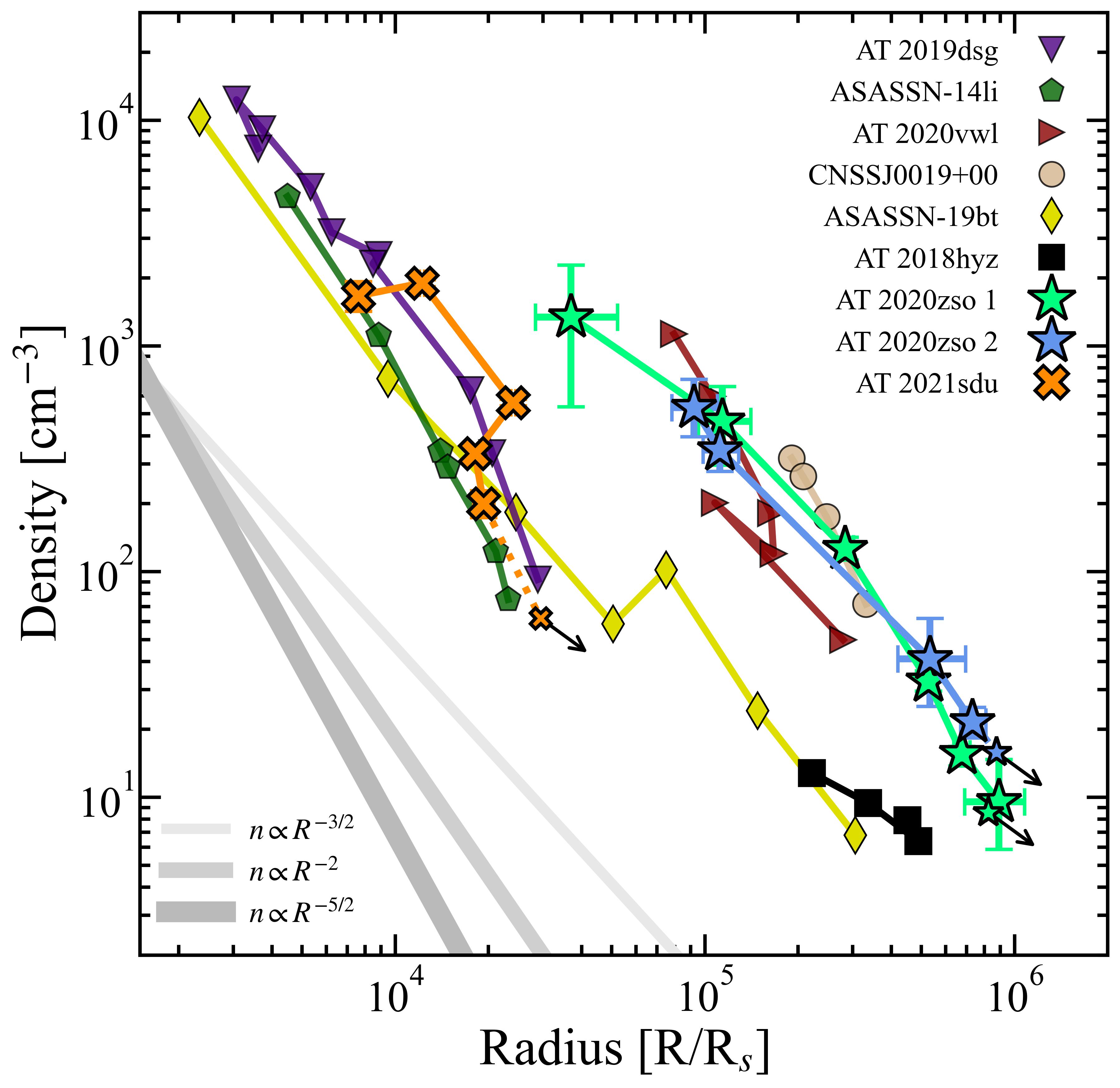}
    \caption{The host galaxy CNM density profile for AT 2020zso and AT 2021sdu in context with other TDE host galaxies. Here we show the radii in terms of the host SMBH Schwarzschild radius ($R_s = 2 G M_{\rm BH}/c^2$) to standardize the comparison. We use the SMBH masses presented in \citet{Wevers_2022} ($M_\text{BH} \approx 10^{6.0}M_\odot$) and \citet{Yao_2023} ($M_\text{BH} \approx 10^{6.7}M_\odot$) for AT 2020zso and AT 2021sdu respectively. The remaining data and assumed SMBH masses are from: ASASSN-14li, \citet{alexander_2016}; CNSS J0019+00, \citet{CNSS}; AT 2019dsg, \citet{at2019dsg}; AT 2018hyz, \cite{AT2018hyz}; AT 2020opy, \citet{at2020opy}; AT 2020vwl, \citet{AT2020vwl}; ASASSN-19bt, \citet{Christy_2024}}
    \label{fig:density}
\end{figure}

In this section, we examine the properties of the CNM surrounding AT 2020zso and AT 2021sdu, and discuss these results in the broader context of the circumnuclear environments in other radio-detected TDEs.

We find evidence for magnetic field inhomogeneities in the CNM surrounding AT 2020zso. At $\delta t \approx 365$ days, the optically thick radio spectrum exhibits a spectral slope of $\alpha = 1.05_{-0.16}^{+0.18}$, significantly softer than the canonical value expected from synchrotron self-absorption ($\alpha = 5/2$; e.g., \citealt{Granot_2002}). This broadening of the radio spectrum can be attributed to a superposition of synchrotron spectra with varying optical depths due to inhomogeneities in the magnetic field strength within the emitting region \citep{bjornsson_2017,Chandra_2019}. By $\delta t \approx 720\text{ d}$, we again capture the optically thick spectral slope and find it consistent with $\alpha = 5/2$, the theoretical expectation for self-absorbed synchrotron emission when $\nu_m < \nu_a$.

This evolution is consistent with a change in CNM conditions near the Bondi radius, $R_B = 2GM_\mathrm{BH}/c_s^2$, where the SMBH's gravitational influence on the ambient gas becomes significant \citep{Rozner2025}. The sound speed in the CNM is given by $c_s = (\gamma k_B T/\mu m_p)^{1/2}$, where $\gamma$ is the adiabatic index, $k_B$ is the Boltzmann constant, $T$ is the CNM temperature, $\mu$ is the mean molecular weight, and $m_p$ is the proton mass. Using a fixed CNM temperature of $T = 10^7$ K, a monatomic gas with $\gamma = 5/3$, and a mean molecular weight $\mu = 0.6$ (following \citealt{Matsumoto_2024}), we estimate $R_B \approx 1.2 \times 10^{17}$ cm for AT 2020zso's SMBH mass ($M_\mathrm{BH} \approx 10^{6.0} M_\odot$; \citealt{Wevers_2022}). The blast wave radius evolves from $R \approx 8.1 \times 10^{16}$ cm at $\delta t \approx 356$ days to $R \approx 1.9 \times 10^{17}$ cm at $\delta t \approx 720$ days, suggesting a transition across $R_B$. We therefore speculate that magnetic field inhomogeneities are more prevalent within $R_B$, potentially due to shocks, turbulence, or prior disk-driven winds in the SMBH's accretion zone \citep[e.g.,][]{Baganoff_2003,Feng_2014}.

The light curve of AT 2020zso closely resembles those of AT 2019dsg and AT 2020vwl, which exhibit delayed radio flares \citep{Cendes_2024,Goodwin_2025}. \citet{Matsumoto_2024} modeled the late-time radio re-brightening in these events as arising from a flattened CNM density profile exterior to the Bondi radius of the central SMBH. While we similarly observe a potential change in the magnetic field structure near $R_B$ in AT 2020zso, we find no corresponding evidence for a change in the CNM density profile itself. 

Figure~\ref{fig:density} shows the inferred CNM density as a function of outflow radius, scaled by the Schwarzschild radius ($R_s = 2GM_\mathrm{BH}/c^2$) of each host SMBH. To facilitate comparison across sources, we recompute all densities assuming the same volume definition we used in our analysis (a spherical shell of thickness $0.1R$). We find that the density profile for AT 2020zso remains consistent for radii above and below $R_B \approx 4.1 \times 10^5 R_s$. Moreover, the density profiles for both flares are in agreement with one another, with $n_{1,\text{ext}}(r)\propto r^{-1.53}$ for the first outflow and $n_{2,\text{ext}}(r)\propto r^{-1.51}$ for the second. Both of these density profiles align with the theoretical expectations for a spherical Bondi accretion history. Given these results, we disfavor the interpretation of density changes near $R_B$ as the origin of the second radio flare in AT 2020zso. 

We note that naively the similarity between the two density profiles is not expected given the geometrical assumptions we have made. For two quasi-spherical outflows, the second outflow should encounter a CNM that was compressed and perturbed by the first outflow. Therefore, the agreement between the two may suggest different geometries between the first and second outflows (e.g., a toroidal vs polar outflow).

For AT 2021sdu, the early SEDs taken at $\delta t \leq 220$ days, particularly at $\delta t \approx 146$ days and $220$ days, show a steeper optically thin spectral index compared to the later epochs. This suggests that the optically thin emission in the early phases originates from a population of cooled electrons (i.e., $\nu_a > \nu_c,\nu_m$). This cooling may be a result of a higher density in the CNM, allowing relativistic electrons to lose energy more efficiently. However, as shown in Figure \ref{fig:density}, the CNM around AT 2021sdu is not particularly over-dense when compared to the host SMBH size. Instead, the outflow appears to be in a similar environment to that of other TDEs, such as ASASSN-14li, AT 2019dsg, and ASASSN-19bt \citep{alexander_2016,at2019dsg, Christy_2024}. This suggests that while the early spectral evolution of AT 2021sdu may reflect local density variations in the CNM, the overall CNM densities are more consistent with those observed in other TDEs.

By integrating the density profiles in Figure \ref{fig:density}, we estimate that the outflows in AT 2020zso have swept up a CNM mass of $M_{\rm swept}\sim7\times10^{-4}M_\odot$, while for AT 2021sdu we find $M_{\rm swept}\sim6\times10^{-5}M_\odot$. For comparison, the kinetic masses implied by the derived blast wave energies and velocities ($M_{\rm k}\approx2E/\beta^2c^2$) are $M_{\rm k}\sim2\times10^{-3}M_\odot$ for both the prompt and delayed outflow in AT 2020zso, and $M_{\rm k}\sim9\times10^{-4}M_\odot$ for AT 2021sdu. The outflows present in both objects are consistent with $M_{\rm kin} > M_{\rm swept}$, which reaffirms the assumption that the blast waves are in free expansion and have not yet entered the Sedov-Taylor phase.

\section{Summary and Conclusions}\label{sec:conc}
We present the discovery and detailed study of the radio transients associated with the TDEs AT 2020zso and AT 2021sdu. We monitored the evolution of the outflows produced by these events using multi-frequency VLA, GMRT, and NOEMA observations. 

We highlight below our key findings for the TDE AT 2020zso:

\begin{itemize}
    \item We find that an energetic outflow was launched shortly after optical peak, and its radio emission reached a maximum at approximately $\delta t \sim 1$ year before declining.

    \item Around $\sim 800$ days post-discovery, the radio light curve of AT 2020zso begins to rise again. Coincident with this, two distinct components are briefly resolved in the radio spectra, suggesting two separate emitting regions and pointing to a physically distinct secondary outflow.

    \item Both outflows present in AT 2020zso share similar properties, with velocities $v \approx 0.1-0.2c$ and energies $E \sim 10^{49}~\mathrm{erg}$ inferred from our equipartition analysis. Additionally, both outflows propagate through a Bondi-like CNM density profile, with possible evidence for magnetic field inhomogeneities within the Bondi radius.

    \item We favor a scenario in which AT 2020zso's first radio flare arises from an accretion-driven outflow launched while the TDE disk was accreting at a relatively high Eddington fraction, consistent with the prompt accretion suggested by \citet{Wevers_2022}. In this picture, as the accretion rate decreases, a second outflow is launched once the disk transitions into a geometrically thick, ADAF-like state.
\end{itemize}

Similarly, our key takeaways for the TDE AT 2021sdu are as follows:

\begin{itemize}
    \item The long-lasting radio emission seen in AT 2021sdu is best explained by a single transient component superimposed on non-variable, resolution-dependent host emission, which becomes dominant around 500 days after discovery and limits the study of its late-time evolution.

    \item Prior to 500 days, the transient radio emission in AT 2021sdu is consistent with a slower outflow ($v \approx 0.03c$), with an inferred launch date about 50 days before optical discovery. This suggests contributions from unbound stellar debris, though accretion-driven disk winds or stream–stream collisions cannot be ruled out.

    \item We observe a rapid decrease in peak flux density between the early and late epochs, which yields an unphysical shock evolution, including a reversal in the outflow radius. This may point to a more complex CNM structure or changes in the outflow structure itself.

    \item The resolution-dependent host contamination highlights an important and sometimes overlooked observational challenge in interpreting late-time radio variability, particularly when using variable-resolution interferometers or combining datasets from multiple facilities.
\end{itemize}

The growing sample of TDEs with delayed radio re-brightenings (e.g., ASASSN-15oi, AT 2020vwl, ASASSN-19bt) suggests that multi-component outflows may reflect an emerging property of the accretion physics involved in TDEs. These findings challenge single-outflow models, favoring instead scenarios where the radio emission in TDEs reflects the evolving accretion processes of the SMBH. 

Our analysis highlights the multi-component nature of the radio emission from AT 2020zso and AT 2021sdu. Disentangling these components was possible thanks to our densely sampled, multi-frequency data, emphasizing the importance of such coverage for accurate interpretation. Future radio monitoring campaigns, combined with other multi-wavelength observations, will be essential for constraining the mechanisms behind both prompt and delayed radio emission.

\section*{Acknowledgments}
The National Radio Astronomy Observatory (NRAO) is a facility of the National Science Foundation operated under cooperative agreement by Associated Universities, Inc. GMRT observations for this study were obtained via projects 43\_046, 44\_022 (PI: Goodwin). We thank the staff of the GMRT who have made these observations possible. This work is based on observations carried out under project numbers E21AA and S22BS with the IRAM NOEMA Interferometer. IRAM is supported by INSU/CNRS (France), MPG (Germany), and IGN (Spain). 

C.~T.~C. and K.~D.~A. acknowledge support provided by the NSF through award SOSPA9-007 from the NRAO and award AST-2307668. N.F. acknowledges support from the National Science Foundation Graduate Research Fellowship Program under Grant No. DGE-2137419. AJG is grateful for support from the Forrest Research Foundation. This work was partially supported by the Australian government through the Australian Research Council’s Discovery Projects funding scheme (DP200102471).  DLC acknowledges support from the Science and Technology Facilities Council (STFC) grant ST/X001121/1. E.~R-R. acknowledges  the Heising-Simons Foundation and NSF: AST 1852393, AST 2150255, and AST 2206243.  Time-domain research by the University of Arizona team and D.J.S. is supported by National Science Foundation (NSF) grants 2108032, 2308181, 2407566, and 2432036 and the Heising-Simons Foundation under grant \#2020-1864.

This work made use of the following software packages: \texttt{astropy} \citep{astropy:2013, astropy:2018, astropy:2022}, \texttt{Jupyter} \citep{2007CSE.....9c..21P, kluyver2016jupyter}, \texttt{matplotlib} \citep{Hunter:2007}, \texttt{numpy} \citep{numpy}, \texttt{pandas} \citep{mckinney-proc-scipy-2010, pandas_16918803}, \texttt{python} \citep{python}, \texttt{scipy} \citep{2020SciPy-NMeth, scipy_17101542}, \texttt{Cython} \citep{cython:2011}, \texttt{emcee} \citep{emcee-Foreman-Mackey-2013, emcee_10996751}, and \texttt{h5py} \citep{collette_python_hdf5_2014, h5py_7560547}.

Software citation information aggregated using \texttt{\href{https://www.tomwagg.com/software-citation-station/}{The Software Citation Station}} \citep{software-citation-station-paper, software-citation-station-zenodo}.

\bibliographystyle{aasjournal}
\bibliography{mybib}

\appendix
\renewcommand{\thetable}{A\arabic{table}}
\setcounter{table}{0}  
\LTcapwidth=\textwidth
\begin{longtable}{clcccccc}
\caption{We present radio observations of AT 2020zso, reporting uncertainties as a 1$\sigma$ statistical error with an additional 5\% systematic error to account for uncertainties in the absolute flux density scale calibration. We report non-detections as 3$\sigma$ upper limits, and we measure all values of $\delta t$ relative to MJD = 59165.7 \citep{20zso_TNS_discovery}.} \\
\label{tab:data_zso}\\
    \hline\hline
    Object & Date & $\delta t$ & Project Code & Array & Configuration & $\nu$ & Flux Density \\ 
     & (UTC) & (d) & & & & (GHz) & ($\mu$Jy) \\[1pt] 
    \hline
    \textbf{AT 2020zso} & 2020 Dec 21 & 40 & 20B-377 & VLA & A & 15.0 & $22 \pm 7 \pm 1$ \\[5pt] 

    & 2021 Feb 20 & 101 & 20B-377 & VLA & A & 9.0 & $63 \pm 16 \pm 3$ \\
    & 2021 Feb 20 & 101 & 20B-377 & VLA & A & 11.0 & $83 \pm 19 \pm 4$ \\
    & 2021 Feb 20 & 101 & 20B-377 & VLA & A & 13.0 & $67 \pm 20 \pm 3$ \\
    & 2021 Feb 20 & 101 & 20B-377 & VLA & A & 15.0 & $63 \pm 18 \pm 3$ \\
    & 2021 Feb 20 & 101 & 20B-377 & VLA & A & 17.0 & $73 \pm 17 \pm 4$ \\[5pt] 
    
    & 2021 May 30 & 198 & ddtC181 & GMRT & - & 0.65 & $<46.59 ^{*}$ \\
    & 2021 Jun 08 & 208 & ddtC181 & GMRT & - & 1.26 & $<51.15 ^{*}$ \\
    & 2021 May 28 & 197 & 20B-377 & VLA & D & 5.0 & $326 \pm 14 \pm 16$ \\
    & 2021 May 28 & 197 & 20B-377 & VLA & D & 7.0 & $275 \pm 13 \pm 14$ \\
    & 2021 May 26 & 195 & 20B-377 & VLA & D & 9.0 & $312 \pm 9 \pm 16$ \\
    & 2021 May 26 & 195 & 20B-377 & VLA & D & 11.0 & $300 \pm 17 \pm 15$ \\
    & 2021 May 26 & 195 & 20B-377 & VLA & D & 12.8 & $248 \pm 21 \pm 12$ \\
    & 2021 May 26 & 195 & 20B-377 & VLA & D & 14.3 & $322 \pm 20 \pm 16$ \\
    & 2021 May 26 & 195 & 20B-377 & VLA & D & 15.9 & $337 \pm 13 \pm 17$ \\
    & 2021 May 26 & 195 & 20B-377 & VLA & D & 17.4 & $299 \pm 14 \pm 15$ \\[5pt] 
    
    & 2021 Nov 2 & 356 & 20B-377 & VLA & B & 1.3 & $269 \pm 85 \pm 13$ \\
    & 2021 Nov 2 & 356 & 20B-377 & VLA & B & 1.8 & $381 \pm 104 \pm 19$ \\
    & 2021 Nov 2 & 356 & 20B-377 & VLA & B & 2.5 & $524 \pm 59 \pm 26$ \\
    & 2021 Nov 2 & 356 & 20B-377 & VLA & B & 3.5 & $622 \pm 27 \pm 31$ \\
    & 2021 Nov 2 & 356 & 20B-377 & VLA & B & 5.0 & $765 \pm 14 \pm 38$ \\
    & 2021 Nov 2 & 356 & 20B-377 & VLA & B & 7.0 & $726 \pm 18 \pm 36$ \\
    & 2021 Nov 2 & 356 & 20B-377 & VLA & B & 9.0 & $625 \pm 14 \pm 31$ \\
    & 2021 Nov 2 & 356 & 20B-377 & VLA & B & 11.0 & $488 \pm 13 \pm 24$ \\
    & 2021 Nov 2 & 356 & 20B-377 & VLA & B & 12.8 & $386 \pm 18 \pm 19$ \\
    & 2021 Nov 2 & 356 & 20B-377 & VLA & B & 14.3 & $339 \pm 14 \pm 17$ \\
    & 2021 Nov 2 & 356 & 20B-377 & VLA & B & 15.9 & $306 \pm 21 \pm 15$ \\
    & 2021 Nov 2 & 356 & 20B-377 & VLA & B & 17.4 & $250 \pm 15 \pm 13$ \\[5pt] 
    
    & 2022 May 12 & 546 & 20B-377 & VLA & A & 1.3 & $489 \pm 46 \pm 24$ \\
    & 2022 May 12 & 546 & 20B-377 & VLA & A & 1.8 & $633 \pm 42 \pm 32$ \\
    & 2022 May 12 & 546 & 20B-377 & VLA & A & 2.5 & $548 \pm 23 \pm 27$ \\
    & 2022 May 12 & 546 & 20B-377 & VLA & A & 3.5 & $504 \pm 18 \pm 25$ \\
    & 2022 May 12 & 546 & 20B-377 & VLA & A & 5.0 & $472 \pm 21 \pm 24$ \\
    & 2022 May 12 & 546 & 20B-377 & VLA & A & 7.0 & $340 \pm 17 \pm 17$ \\
    & 2022 May 12 & 546 & 20B-377 & VLA & A & 9.0 & $274 \pm 12 \pm 14$ \\
    & 2022 May 12 & 546 & 20B-377 & VLA & A & 11.0 & $219 \pm 19 \pm 11$ \\
    & 2022 May 12 & 546 & 20B-377 & VLA & A & 13.0 & $152 \pm 14 \pm 8$ \\
    & 2022 May 12 & 546 & 20B-377 & VLA & A & 15.0 & $149 \pm 16 \pm 7$ \\
    & 2022 May 12 & 546 & 20B-377 & VLA & A & 17.0 & $143 \pm 20 \pm 7$ \\[5pt]
    
    & 2022 Oct 10 & 717 & 43\_046 & GMRT & - & 0.65 & $197 \pm 31 \pm 10$ \\
    & 2022 Nov 2 & 720 & 20B-377 & VLA & C & 1.3 & $506 \pm 109 \pm 25$ \\
    & 2022 Nov 2 & 720 & 20B-377 & VLA & C & 1.8 & $422 \pm 56 \pm 21$ \\
    & 2022 Nov 2 & 720 & 20B-377 & VLA & C & 2.5 & $448 \pm 51 \pm 22$ \\
    & 2022 Nov 2 & 720 & 20B-377 & VLA & C & 3.5 & $333 \pm 31 \pm 17$ \\
    & 2022 Nov 2 & 720 & 20B-377 & VLA & C & 5.0 & $213 \pm 21 \pm 11$ \\
    & 2022 Nov 2 & 720 & 20B-377 & VLA & C & 7.0 & $172 \pm 19 \pm 9$ \\
    & 2022 Nov 2 & 720 & 20B-377 & VLA & C & 9.0 & $118 \pm 11 \pm 6$ \\
    & 2022 Nov 2 & 720 & 20B-377 & VLA & C & 11.0 & $116 \pm 16 \pm 6$ \\
    & 2022 Nov 2 & 720 & 20B-377 & VLA & C & 12.8 & $74 \pm 17 \pm 4$ \\
    & 2022 Nov 2 & 720 & 20B-377 & VLA & C & 14.3 & $54 \pm 14 \pm 3$ \\
    & 2022 Nov 2 & 720 & 20B-377 & VLA & C & 15.9 & $46 \pm 16 \pm 2$ \\
    & 2022 Nov 2 & 720 & 20B-377 & VLA & C & 17.4 & $41 \pm 18 \pm 2$ \\[5pt] 
    
    & 2023 Apr 13 & 882 & 20B-377 & VLA & B & 1.3 & $499 \pm 48 \pm 25$ \\
    & 2023 Apr 13 & 882 & 20B-377 & VLA & B & 1.8 & $463 \pm 41 \pm 23$ \\
    & 2023 Apr 13 & 882 & 20B-377 & VLA & B & 2.5 & $329 \pm 94 \pm 16$ \\
    & 2023 Apr 13 & 882 & 20B-377 & VLA & B & 3.5 & $306 \pm 22 \pm 15$ \\
    & 2023 Apr 13 & 882 & 20B-377 & VLA & B & 5.0 & $287 \pm 21 \pm 14$ \\
    & 2023 Apr 13 & 882 & 20B-377 & VLA & B & 7.0 & $252 \pm 13 \pm 13$ \\
    & 2023 Apr 13 & 882 & 20B-377 & VLA & B & 9.0 & $215 \pm 14 \pm 11$ \\
    & 2023 Apr 13 & 882 & 20B-377 & VLA & B & 11.0 & $172 \pm 18 \pm 9$ \\[5pt] 
    
    & 2023 May 27 & 926 & 20B-377 & VLA & B & 1.3 & $436 \pm 55 \pm 22$ \\
    & 2023 May 27 & 926 & 20B-377 & VLA & B & 1.8 & $287 \pm 57 \pm 14$ \\
    & 2023 May 27 & 926 & 20B-377 & VLA & B & 2.5 & $234 \pm 35 \pm 12$ \\
    & 2023 May 27 & 926 & 20B-377 & VLA & B & 3.5 & $255 \pm 16 \pm 13$ \\
    & 2023 May 27 & 926 & 20B-377 & VLA & B & 5.0 & $248 \pm 14 \pm 12$ \\
    & 2023 May 27 & 926 & 20B-377 & VLA & B & 7.0 & $192 \pm 10 \pm 10$ \\
    & 2023 May 27 & 926 & 20B-377 & VLA & B & 9.0 & $160 \pm 10 \pm 8$ \\
    & 2023 May 27 & 926 & 20B-377 & VLA & B & 11.0 & $130 \pm 12 \pm 6$ \\
    & 2023 May 27 & 926 & 20B-377 & VLA & B & 12.8 & $88 \pm 22 \pm 4$ \\
    & 2023 May 27 & 926 & 20B-377 & VLA & B & 14.3 & $88 \pm 11 \pm 4$ \\
    & 2023 May 27 & 926 & 20B-377 & VLA & B & 15.9 & $75 \pm 16 \pm 4$ \\
    & 2023 May 27 & 926 & 20B-377 & VLA & B & 17.4 & $80 \pm 17 \pm 4$ \\[5pt] 
    
    & 2023 Oct 1 & 1053 & 20B-377 & VLA & A & 2.5 & $446 \pm 42 \pm 22$ \\
    & 2023 Oct 1 & 1053 & 20B-377 & VLA & A & 3.5 & $370 \pm 30 \pm 19$ \\
    & 2023 Oct 1 & 1053 & 20B-377 & VLA & A & 5.0 & $286 \pm 10 \pm 14$ \\
    & 2023 Oct 1 & 1053 & 20B-377 & VLA & A & 7.0 & $175 \pm 8 \pm 9$ \\
    & 2023 Oct 1 & 1053 & 20B-377 & VLA & A & 9.0 & $137 \pm 9 \pm 7$ \\
    & 2023 Oct 1 & 1053 & 20B-377 & VLA & A & 11.0 & $84 \pm 15 \pm 4$ \\
    & 2023 Oct 1 & 1053 & 20B-377 & VLA & A & 13.5 & $81 \pm 18 \pm 4$ \\
    & 2023 Oct 1 & 1053 & 20B-377 & VLA & A & 16.5 & $68 \pm 8 \pm 3$ \\[5pt] 
    
    & 2024 May 9 & 1275 & 20B-377 & VLA & B & 1.3 & $498 \pm 44 \pm 25$ \\
    & 2024 May 9 & 1275 & 20B-377 & VLA & B & 1.8 & $467 \pm 60 \pm 23$ \\
    & 2024 May 9 & 1275 & 20B-377 & VLA & B & 2.5 & $347 \pm 21 \pm 17$ \\
    & 2024 May 9 & 1275 & 20B-377 & VLA & B & 3.5 & $278 \pm 19 \pm 14$ \\
    & 2024 May 9 & 1275 & 20B-377 & VLA & B & 5.0 & $187 \pm 8 \pm 9$ \\
    & 2024 May 9 & 1275 & 20B-377 & VLA & B & 7.0 & $133 \pm 8 \pm 7$ \\
    & 2024 May 9 & 1275 & 20B-377 & VLA & B & 9.0 & $92 \pm 7 \pm 5$ \\
    & 2024 May 9 & 1275 & 20B-377 & VLA & B & 11.0 & $74 \pm 8 \pm 4$ \\
    & 2024 May 9 & 1275 & 20B-377 & VLA & B & 13.5 & $54 \pm 6 \pm 3$ \\
    & 2024 May 9 & 1275 & 20B-377 & VLA & B & 16.5 & $40 \pm 7 \pm 2$ \\[5pt] 
    
    & 2024 Aug 25 & 1382 & 20B-377 & VLA & B & 1.3 & $650 \pm 35 \pm 32$ \\
    & 2024 Aug 25 & 1382 & 20B-377 & VLA & B & 1.8 & $569 \pm 66 \pm 28$ \\
    & 2024 Aug 25 & 1382 & 20B-377 & VLA & B & 2.5 & $334 \pm 15 \pm 17$ \\
    & 2024 Aug 25 & 1382 & 20B-377 & VLA & B & 3.5 & $215 \pm 13 \pm 11$ \\
    & 2024 Aug 25 & 1382 & 20B-377 & VLA & B & 5.0 & $194 \pm 14 \pm 10$ \\
    & 2024 Aug 25 & 1382 & 20B-377 & VLA & B & 7.0 & $110 \pm 8 \pm 6$ \\
    & 2024 Aug 25 & 1382 & 20B-377 & VLA & B & 9.0 & $81 \pm 6 \pm 4$ \\
    & 2024 Aug 25 & 1382 & 20B-377 & VLA & B & 11.0 & $70 \pm 11 \pm 4$ \\
    & 2024 Aug 25 & 1382 & 20B-377 & VLA & B & 13.5 & $52 \pm 6 \pm 3$ \\
    & 2024 Aug 25 & 1382 & 20B-377 & VLA & B & 16.5 & $42 \pm 10 \pm 2$ \\[5pt] 

    \hline\hline
\end{longtable}
\noindent\textit{Note:} $^*$ Data taken from \citet{Roy_2021}.

\begin{longtable}{clcccccc}
\caption{We present radio observations of AT 2021sdu, reporting uncertainties as a 1$\sigma$ statistical error with an additional 5\% systematic error to account for uncertainties in the absolute flux density scale calibration. We report non-detections as 3$\sigma$ upper limits, and we measure all values of $\delta t$ relative to MJD = 59400.9 \citep{21sdu_TNS_discovery}.}\\

    \hline\hline
    Object & Date & $\delta t$ & Project Code & Array & Configuration & $\nu$ & Flux Density \\ 
     & (UTC) & (d) & & & & (GHz) & ($\mu$Jy) \\[1pt] 
    \hline
    \textbf{AT 2021sdu}& 2021 Aug 10 & 36 & 20B-377 & VLA & C & 15.1 & $48 \pm 7 \pm 2$ \\[5pt] 

    & 2021 Sep 11 & 68 & 20B-377 & VLA & C & 9.0 & $165 \pm 13 \pm 8$ \\
    & 2021 Sep 11 & 68 & 20B-377 & VLA & C & 11.0 & $156 \pm 9 \pm 8$ \\
    & 2021 Sep 11 & 68 & 20B-377 & VLA & C & 12.8 & $161 \pm 11 \pm 8$ \\
    & 2021 Sep 11 & 68 & 20B-377 & VLA & C & 14.3 & $160 \pm 9 \pm 8$ \\
    & 2021 Sep 11 & 68 & 20B-377 & VLA & C & 15.8 & $174 \pm 12 \pm 9$ \\
    & 2021 Sep 11 & 68 & 20B-377 & VLA & C & 17.4 & $197 \pm 12 \pm 10$ \\[5pt] 
    
    & 2021 Nov 28 & 146 & 20B-377 & VLA & B & 5.0 & $84 \pm 10 \pm 4$ \\
    & 2021 Nov 28 & 146 & 20B-377 & VLA & B & 7.0 & $170 \pm 12 \pm 8$ \\
    & 2021 Nov 28 & 146 & 20B-377 & VLA & B & 9.0 & $299 \pm 12 \pm 15$ \\
    & 2021 Nov 28 & 146 & 20B-377 & VLA & B & 11.0 & $478 \pm 15 \pm 24$ \\
    & 2021 Nov 28 & 146 & 20B-377 & VLA & B & 12.8 & $628 \pm 14 \pm 31$ \\
    & 2021 Nov 28 & 146 & 20B-377 & VLA & B & 14.3 & $667 \pm 16 \pm 33$ \\
    & 2021 Nov 28 & 146 & 20B-377 & VLA & B & 15.9 & $690 \pm 16 \pm 35$ \\
    & 2021 Nov 28 & 146 & 20B-377 & VLA & B & 17.4 & $690 \pm 13 \pm 35$ \\
    & 2021 Nov 28 & 146 & 20B-377 & VLA & B & 19.1 & $736 \pm 159 \pm 37$ \\
    & 2021 Nov 28 & 146 & 20B-377 & VLA & B & 21.0 & $680 \pm 28 \pm 34$ \\
    & 2021 Nov 28 & 146 & 20B-377 & VLA & B & 23.0 & $646 \pm 17 \pm 32$ \\
    & 2021 Nov 28 & 146 & 20B-377 & VLA & B & 24.9 & $615 \pm 16 \pm 31$ \\
    & 2021 Dec 21 & 169 & E21AA & NOEMA & - & 88.5 & $138 \pm 15 \pm 7$ \\[5pt] 
    
    & 2022 Jan 26 & 205 & 20B-377 & VLA & B & 3.0 & $185 \pm 18 \pm 9$ \\
    & 2022 Jan 26 & 205 & 20B-377 & VLA & B & 5.0 & $456 \pm 13 \pm 23$ \\
    & 2022 Jan 26 & 205 & 20B-377 & VLA & B & 7.0 & $650 \pm 13 \pm 33$ \\
    & 2022 Jan 26 & 205 & 20B-377 & VLA & B & 9.0 & $743 \pm 16 \pm 37$ \\
    & 2022 Jan 26 & 205 & 20B-377 & VLA & B & 11.0 & $804 \pm 21 \pm 40$ \\
    & 2022 Jan 26 & 205 & 20B-377 & VLA & B & 12.8 & $804 \pm 18 \pm 40$ \\
    & 2022 Jan 26 & 205 & 20B-377 & VLA & B & 14.3 & $755 \pm 22 \pm 38$ \\
    & 2022 Jan 26 & 205 & 20B-377 & VLA & B & 15.9 & $715 \pm 15 \pm 36$ \\
    & 2022 Jan 26 & 205 & 20B-377 & VLA & B & 17.4 & $683 \pm 19 \pm 34$ \\
    & 2022 Jan 26 & 205 & 20B-377 & VLA & B & 19.1 & $604 \pm 14 \pm 30$ \\
    & 2022 Jan 26 & 205 & 20B-377 & VLA & B & 21.0 & $532 \pm 27 \pm 27$ \\
    & 2022 Jan 26 & 205 & 20B-377 & VLA & B & 23.0 & $484 \pm 19 \pm 24$ \\
    & 2022 Jan 26 & 205 & 20B-377 & VLA & B & 24.9 & $462 \pm 23 \pm 23$ \\
    & 2022 Feb 10 & 220 & 20B-377 & VLA & BnA & 31.0 & $304 \pm 17 \pm 15$ \\
    & 2022 Feb 10 & 220 & 20B-377 & VLA & BnA & 35.0 & $231 \pm 16 \pm 12$ \\
    & 2022 Feb 10 & 220 & 20B-377 & VLA & BnA & 42.0 & $201 \pm 24 \pm 10$ \\
    & 2022 Feb 10 & 220 & 20B-377 & VLA & BnA & 46.0 & $196 \pm 41 \pm 10$ \\
    & 2022 Jan 26 & 205 & E21AA & NOEMA & - & 88.5 & $105 \pm 20 \pm 5$ \\[5pt] 
    
    & 2022 May 15 & 314 & 20B-377 & VLA & A & 2.5 & $131 \pm 18 \pm 7$ \\
    & 2022 May 15 & 314 & 20B-377 & VLA & A & 3.5 & $147 \pm 12 \pm 7$ \\
    & 2022 May 15 & 314 & 20B-377 & VLA & A & 5.0 & $181 \pm 14 \pm 9$ \\
    & 2022 May 15 & 314 & 20B-377 & VLA & A & 7.0 & $250 \pm 13 \pm 12$ \\
    & 2022 May 15 & 314 & 20B-377 & VLA & A & 9.0 & $228 \pm 13 \pm 11$ \\
    & 2022 May 15 & 314 & 20B-377 & VLA & A & 11.0 & $203 \pm 18 \pm 10$ \\
    & 2022 May 15 & 314 & 20B-377 & VLA & A & 13.0 & $208 \pm 13 \pm 10$ \\
    & 2022 May 15 & 314 & 20B-377 & VLA & A & 15.1 & $181 \pm 13 \pm 9$ \\
    & 2022 May 15 & 314 & 20B-377 & VLA & A & 17.1 & $163 \pm 14 \pm 8$ \\
    & 2022 May 15 & 314 & 20B-377 & VLA & A & 19.0 & $198 \pm 16 \pm 10$ \\
    & 2022 May 15 & 314 & 20B-377 & VLA & A & 21.0 & $132 \pm 15 \pm 7$ \\
    & 2022 May 15 & 314 & 20B-377 & VLA & A & 23.0 & $123 \pm 17 \pm 6$ \\
    & 2022 May 15 & 314 & 20B-377 & VLA & A & 25.0 & $126 \pm 16 \pm 6$ \\
    & 2022 Apr 12 & 281 & E21AA & NOEMA & - & 88.5 & $<47$ \\[5pt] 
    
    & 2022 Jun 19 & 349 & 20B-377 & VLA & A & 1.5 & $132 \pm 38 \pm 7$ \\
    & 2022 Jun 19 & 349 & 20B-377 & VLA & A & 2.5 & $137 \pm 27 \pm 7$ \\
    & 2022 Jun 19 & 349 & 20B-377 & VLA & A & 3.5 & $164 \pm 14 \pm 8$ \\
    & 2022 Jun 19 & 349 & 20B-377 & VLA & A & 5.0 & $166 \pm 18 \pm 8$ \\
    & 2022 Jun 19 & 349 & 20B-377 & VLA & A & 7.0 & $133 \pm 16 \pm 7$ \\
    & 2022 Jun 19 & 349 & 20B-377 & VLA & A & 9.0 & $134 \pm 21 \pm 7$ \\
    & 2022 Jun 19 & 349 & 20B-377 & VLA & A & 11.0 & $122 \pm 28 \pm 6$ \\
    & 2022 Jun 19 & 349 & 20B-377 & VLA & A & 15.1 & $90 \pm 16 \pm 5$ \\
    & 2022 Jun 19 & 349 & 20B-377 & VLA & A & 22.0 & $61 \pm 19 \pm 3$ \\
    & 2022 Jul 07 & 367 & S22BS & NOEMA & - & 88.5 & $<48$ \\[5pt] 
    
    & 2022 Oct 9 & 461 & 20B-377 & VLA & C & 2.5 & $315 \pm 34 \pm 16$ \\
    & 2022 Oct 9 & 461 & 20B-377 & VLA & C & 3.5 & $211 \pm 27 \pm 11$ \\
    & 2022 Oct 9 & 461 & 20B-377 & VLA & C & 5.0 & $139 \pm 25 \pm 7$ \\
    & 2022 Oct 9 & 461 & 20B-377 & VLA & C & 7.0 & $100 \pm 19 \pm 5$ \\
    & 2022 Oct 9 & 461 & 20B-377 & VLA & C & 10.0 & $67 \pm 11 \pm 3$ \\
    & 2022 Oct 9 & 461 & 20B-377 & VLA & C & 15.1 & $58 \pm 12 \pm 3$ \\
    & 2022 Sep 30 & 452 & S22BS & NOEMA & - & 88.5 & $<57$ \\[5pt] 
    
    & 2023 Apr 22 & 656 & 44\_022 & GMRT & - & 0.65 & $566 \pm 91 \pm 28$ \\
    & 2023 Apr 21 & 655 & 44\_022 & GMRT & - & 1.26 & $260 \pm 51 \pm 13$ \\
    & 2023 May 13 & 677 & 20B-377 & VLA & B & 1.5 & $162 \pm 47 \pm 8$ \\
    & 2023 May 13 & 677 & 20B-377 & VLA & B & 3.0 & $83 \pm 26 \pm 4$ \\
    & 2023 May 13 & 677 & 20B-377 & VLA & B & 6.0 & $51 \pm 12 \pm 3$ \\
    & 2023 May 13 & 677 & 20B-377 & VLA & B & 10.0 & $39 \pm 13 \pm 2$ \\
    & 2023 May 13 & 677 & 20B-377 & VLA & B & 15.1 & $29 \pm 8 \pm 1$ \\[5pt] 
    
    & 2023 Sep 05 & 792 & 44\_022 & GMRT & - & 0.65 & $699 \pm 97 \pm 35$ \\
    & 2023 Sep 04 & 791 & 44\_022 & GMRT & - & 1.26 & $222 \pm 40 \pm 11$ \\
    & 2023 Sep 30 & 817 & 20B-377 & VLA & A & 1.5 & $151 \pm 28 \pm 8$ \\
    & 2023 Sep 30 & 817 & 20B-377 & VLA & A & 3.0 & $63 \pm 18 \pm 3$ \\
    & 2023 Sep 30 & 817 & 20B-377 & VLA & A & 6.0 & $46 \pm 11 \pm 2$ \\
    & 2023 Sep 30 & 817 & 20B-377 & VLA & A & 10.0 & $<30$ \\[5pt] 
    
    & 2024 Jan 15 & 965 & 45\_123 & GMRT & - & 0.65 & $425 \pm 108 \pm 21$ \\
    & 2024 Jan 15 & 965 & 45\_123 & GMRT & - & 1.26 & $270 \pm 59 \pm 14$ \\
    & 2024 May 10 & 1040 & 20B-377 & VLA & B & 1.8 & $175 \pm 27 \pm 9$ \\
    & 2024 May 10 & 1040 & 20B-377 & VLA & B & 3.0 & $88 \pm 19 \pm 4$ \\
    & 2024 May 10 & 1040 & 20B-377 & VLA & B & 6.0 & $<24$ \\
    & 2024 May 10 & 1040 & 20B-377 & VLA & B & 10.0 & $<16$ \\[5pt] 
    
    & 2024 Aug 25 & 1147 & 20B-377 & VLA & B & 1.5 & $209 \pm 29 \pm 10$ \\
    & 2024 Aug 25 & 1147 & 20B-377 & VLA & B & 3.0 & $62 \pm 15 \pm 3$ \\
    & 2024 Aug 25 & 1147 & 20B-377 & VLA & B & 6.0 & $32 \pm 8 \pm 2$ \\
    & 2024 Aug 25 & 1147 & 20B-377 & VLA & B & 10.0 & $<21$ \\

    \hline\hline
    \label{tab:data_sdu}
    
\end{longtable}


\label{lastpage}
\end{document}